%% file: main_preprint.tex
\pdfoutput=1
\documentclass[reqno]{amsart}
\usepackage{algorithm}
\usepackage{amsmath, amssymb}
\usepackage[foot]{amsaddr}
\usepackage{bm}
\usepackage{geometry}
\usepackage{graphicx}
\usepackage{hyperref}
\usepackage[utf8]{inputenc}
\usepackage{lineno}
\usepackage{mathtools}
\usepackage{pdfpages}
\usepackage{url} 
\usepackage{xcolor}
\makeatletter
\renewcommand{\email}[2][]{%
	\ifx\emails\@empty\relax\else{\g@addto@macro\emails{,\space}}\fi%
	\@ifnotempty{#1}{\g@addto@macro\emails{\textrm{(#1)}\space}}%
	\g@addto@macro\emails{#2}%
}
\makeatother
\newcommand{\citeA}{\cite}

\title{Bayesian calibration and sensitivity analysis for a karst aquifer model using active subspaces}
\author[M. Teixeira Parente]{Mario Teixeira Parente$^\dagger{}^\P$}
\author[D. Bittner]{Daniel Bittner$^*$}
\author[S. Mattis]{Steven Mattis$^\dagger$}
\author[G. Chiogna]{Gabriele Chiogna$^*{^+}$}
\author[B. Wohlmuth]{Barbara Wohlmuth$^\dagger{}$}
\address{$^\dagger$ Chair for Numerical Mathematics, Technical University Munich, Germany}
\address{$^*$ Chair of Hydrology and River Basin Management, Technical University Munich, Germany}
\address{$^+$ Institute of Geography, University of Innsbruck, Austria}
\address{$^\P$ \textnormal{Corresponding author}}
\email{$\lbrace \text{parente,mattis,wohlmuth}\rbrace$@ma.tum.de}
\email{$\lbrace \text{daniel.bittner,gabriele.chiogna}\rbrace$@tum.de}
\date{\today}
\begin{document}
\input{abstract.tex}
\maketitle
\input{introduction.tex}
\input{model_data.tex}
\input{inference.tex}
\input{results.tex}
\input{discussion.tex}
\input{summary.tex}
\appendix
\input{appendix.tex}
\section*{Acknowledgments}
\input{acknowledgments.tex}
\bibliographystyle{plain}

\input{main_preprint.bbl}
\includepdf[pages=-]{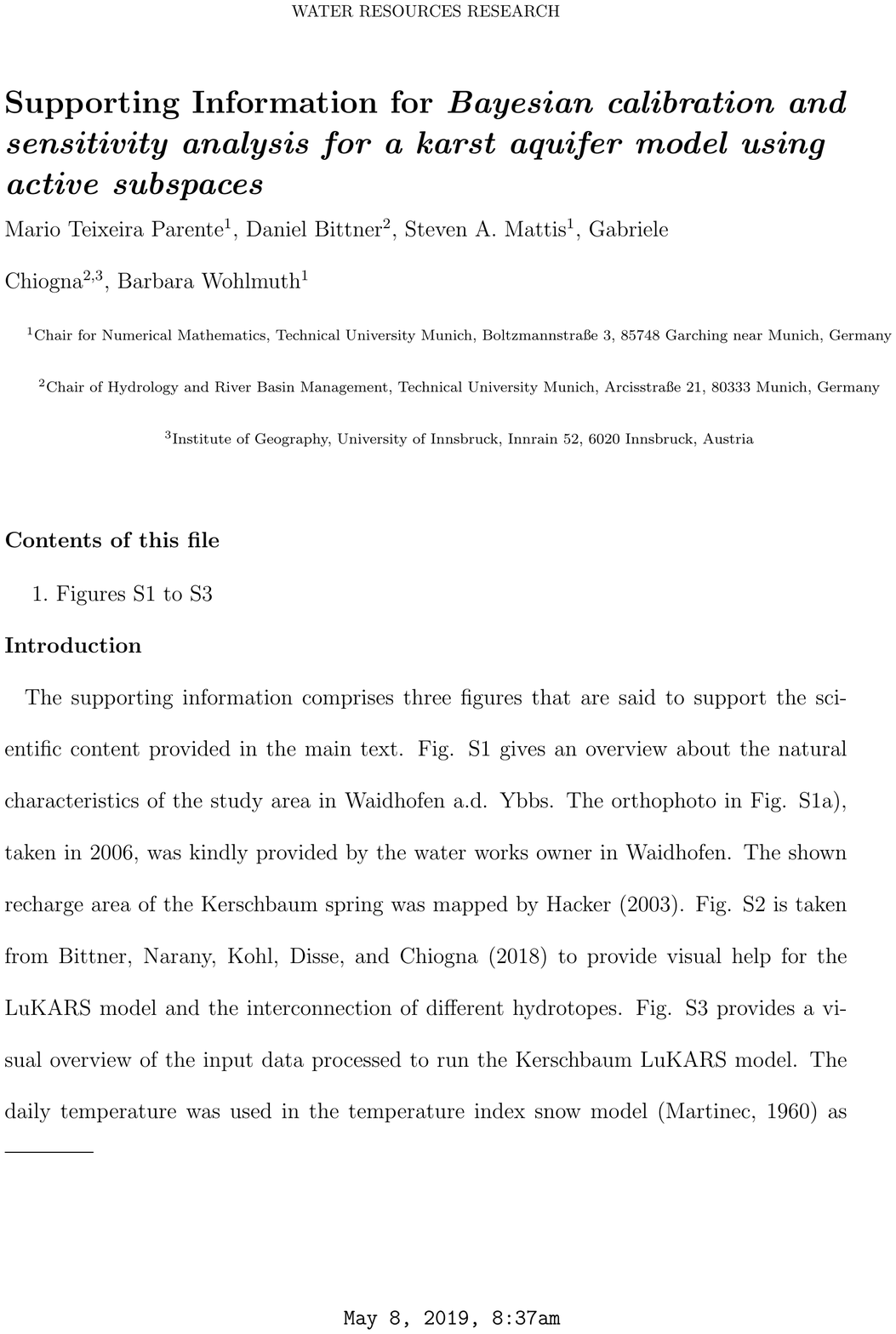}
\end{document}

%% file: abstract.tex
\begin{abstract}
In this article, we perform a parameter study for a recently developed karst hydrological model.
The study consists of a high-dimensional Bayesian inverse problem and a global sensitivity analysis.
For the first time in karst hydrology, we use the active subspace method to find directions in the parameter space that dominate the Bayesian update from the prior to the posterior distribution in order to effectively reduce the dimension of the problem and for computational efficiency.
Additionally, the calculated active subspace can be exploited to construct sensitivity metrics on each of the individual parameters and be used to construct a natural model surrogate.
The model consists of 21 parameters to reproduce the hydrological behavior of spring discharge in a karst aquifer located in the Kerschbaum spring recharge area at Waidhofen a.d. Ybbs in Austria.
The experimental spatial and time series data for the inference process were collected by the water works in Waidhofen.
We show that this case study has implicit low-dimensionality, and we run an adjusted Markov chain Monte Carlo algorithm in a low-dimensional subspace to construct samples of the posterior distribution.
The results are visualized and verified by plots of the posterior's push-forward distribution displaying the uncertainty in predicting discharge values due to the experimental noise in the data.
Finally, a discussion provides hydrological interpretation of these results for the Kerschbaum area.
\end{abstract}

%% file: introduction.tex
\section{Introduction}
\label{sec:introduction}

Models are commonly used in karst systems to investigate the dominant hydrological processes and the quantity and quality of water resources in well-defined surface or subsurface catchments.
Various karst modeling approaches exist, ranging from black-box models \cite{labat2000rainfall,labat1999linear,jukic2008estimating}, i.\,e., transferring an input signal to a desired output signal, over lumped parameter models (grey-box) \cite{fleury2009modelling,mazzilli2017karstmod,sivelle2019dynamics} to distributed process-based models \cite{reimann2009modflow,giese2018turbulent,Henson.2018,sauter2006modellierung}. 
Given their ability to represent the physical characteristics of a catchment in detail, distributed process-based models are usually the first choice in water resources research.
In the particular case of karst aquifers, however, acquiring the relevant data for these models is challenging due to the heterogeneous nature of karstic systems and their mostly unknown subsurface drainage systems \cite{xu2018characterization}.
Also, past studies have shown that even if physical parameters may be obtained from field observations, the fact that they mostly represent point measurements can lead to a severe mismatch when using these parameters in distributed hydrological models \cite{hogue2006evaluating,rosero2010quantifying}. 
For these reasons, lumped process-based models are commonly accepted modeling approaches in karst water resources research \cite{jukic2009groundwater,jourde2015karstmod,hartmann2013process}.
The parameters of such lumped modeling approaches are typically not directly measurable in the field and need to be estimated in the framework of model calibration \cite{hartmann2017value}.
This leads to a decisive trade-off: on the one hand, lumped models based on a low number of calibration parameters, e.\,g., 4 to 6, are less prone to non-uniqueness in parameter identification \cite{jakeman1993much,beven2006manifesto}, i.\,e., different parameter combinations lead to the same result.
However, the representation of the dominant hydrological processes in karst systems may be too simple and not sufficiently represented by this low number of parameters \cite{hartmann2013testing}.
In contrast, by including more calibration parameters to better represent relevant processes in the model structure, such as the effect of land use changes on spring discharges, the parameters may become unidentifiable, which can reduce the prediction accuracy of the model \cite{hartmann2014karst}.
To tackle the challenge of applying lumped parameter models with a high-dimensional parameter space for karst hydrological research studies, there is a need to perform comprehensive parameter studies to avoid model overparametrization and to reduce model parameter and output uncertainties.

With the rise of computational power in the last two decades, Bayesian inverse problems have become a popular part of comprehensive parameter studies for hydrological models \cite{engeland2002bayesian,kavetski2006bayesian,kitanidis2014principal,thiemann2001bayesian}.
In contrast to classical inverse problems, whose formulation is often ill-posed, a Bayesian problem formulation introduces regularizing prior information which gives a different formulation that is mathematically well behaved.
Bayesian inversion aims at finding a posterior distribution on the model parameters incorporating the information of measured, noisy data, e.\,g., spring discharge.
The posterior, which is also used to quantify uncertainty in the inferred parameters, is proportional to the product of the likelihood function and the prior, a distribution that is assumed to be known for the model parameters based on prior knowledge, e.\,g., resulting from field campaigns or manual model calibration.

The most common strategy used for approximating a posterior is to construct samples from the distribution with desired properties.
\textit{Markov chain Monte Carlo} (MCMC) and its derivatives are popular sample-based techniques \cite{vrugt2008treatment,martin2020simplified}; however, they present computational challenges.
These algorithms generally require a very large number of evaluations of the model to provide an acceptable result.
In the hydrological community, where a single model evaluation is often quite computationally expensive, naive implementations of these methods may not be viable, and having too few samples causes the solution to be polluted with sampling error.
One way to reduce this effect is to construct a surrogate model that has a much lower computational cost.
Common approaches for global surrogate models include stochastic finite element approaches for forming polynomial approximations \cite{XK_PC_2002, WK_PC_2006, LeMaitre_UQ_04a, LeMaitre_UQ_04b} and using tensor grid or sparse grid stochastic collocation methods \cite{Almeida_Oden_2010, NTW_SparseGrid_2008}.
Another strategy is to use adaptive sampling methods to reduce sample-based error as shown in e.\,g., \citeA{haario2006dram,vrugt2009accelerating,mattis2018goal}.
Another attribute that affects the computational expense of sample-based approximation methods is high-dimensionality in the space of uncertain parameters because it may slow the identification of areas of high probability and has the tendency to produce highly correlated samples.
There has been much effort in developing algorithms that are more efficient and reduce computational expense due to high-dimensional parameter spaces by effectively reducing the dimension \cite{boyce2014parameter,bui2014solving,constantine2016accelerating,cui2016dimension}.

A relatively new technique for dimension reduction is the \textit{active subspace method} presented in \citeA{constantine2014active}, \citeA{constantine2015active}, and \citeA{russi2010uncertainty}, which seeks orthogonal directions in the space of parameters that dominate the Bayesian update from the prior to the posterior distribution.
These dominant directions span the subspace and define a new coordinate system a low-dimensional Markov chain can move in.
Chains in lower dimensions have preferable properties concerning autocorrelation times which makes them more efficient when producing posterior samples of hydrological model parameters.
As a bonus, the active subspace can be exploited also to calculate global sensitivity metrics for individual parameters as shown in \citeA{constantine2017global}.
The active subspace method has previously been successful in reducing the effective dimension of parameter spaces, e.\,g., for efficient Bayesian inversion of a complex subsurface process \cite{teixeiraparente2019efficient}, or to study sensitivities in a hydrological model \cite{jefferson2015active}.
An added benefit of the active subspace method is that there is a natural cheap global surrogate model embedded in the method via global polynomial regression in the low-dimensional active subspace.
Thus the advantage of the active subspace method is threefold: it effectively reduces the dimension of the Bayesian inverse problem, it easily produces global sensitivity metrics, and it naturally allows for the construction of a cheap global surrogate model.
All of these are gained from the same moderate number of forward model evaluations.

While several hydrological studies address the issue of model parameter sensitivities for inference (e.\,g. \citeA{cuntz2016impact,mockler2016understanding,vanuytrecht2014global}), there has been little effort in karst hydrological research to investigate the low-dimen\-sionality of a corresponding parameter estimation problem (e.\,g., \citeA{sarrazin2018v2karst}).
Our approach is different to these studies in the sense that we study dominant \textit{directions} in parameter spaces and do not focus on sensitivities of coordinate-aligned, i.\,e., individual parameters.
In this regard, the objective of this manuscript is the introduction of the active subspace method to the field of karst hydrology as a technique for dimension reduction and sensitivity analysis in parameter studies.
We demonstrate this method and its mentioned advantages by investigating parameter relationships in the LuKARS model, a lumped karst aquifer model with a high-dimensional parameter space that was recently developed by \citeA{bittner2018modeling} to perform land use change impact studies in karstic environments.
We hypothesize that it is possible to reduce the dimensions of the parameter space in LuKARS, thus saving computational cost, and to better constrain the parameter ranges of the most sensitive model parameters leading to a reduction in model parameter and model result uncertainties.

This article is organized as follows.
Section~\ref{sec:model_data} provides a brief introduction into the study area and the structure of the LuKARS model.
In Section~\ref{sec:inference}, we explain Bayesian inversion, how we exploit active subspaces for it, the construction of global sensitivity values and the concrete setting for our application.
The computational results are presented in Section~\ref{sec:results}, followed by a comprehensive discussion in Section~\ref{sec:discussion} in which we also comment on limits and transferability of the proposed method.
Finally, we conclude with a summary in Section~\ref{sec:summary}.

%% file: model_data.tex
\section{Case study}
\label{sec:model_data}
\subsection{Kerschbaum spring recharge area}
The karst spring that we investigate in the present study is the Kerschbaum spring located about 10km south of the city of Waidhofen a.d. Ybbs (Austria) (Fig~S1~a and b).
Its recharge area was delimited in a former study by \citeA{Hacker.2003} and comprises about 2.5\,km$^2$.
This pre-alpine catchment is part of the eastern-most foothills of the Northern Calcareous Alps with the lowest elevation of 435\,m at the Kerschbaum spring and a maximum elevation of 868\,m on the summit of the mountain Glash\"uttenberg.
The climate of the study area can be described as warm-moderate, with an annual mean temperature of 8$^\circ$\,C and an annual mean precipitation of 1379\,mm, both determined from daily measuring data recorded at the Hinterlug weather station between 1981 and 2014.
Forests represent the dominant land cover in the study area with beeches as primary tree species.
Moreover, parts of the recharge area are used for dolomite mining.

From a geological point of view, the entire recharge area of the Kerschbaum spring is dominated by a lithologic sequence of Triassic dolostones (Fig.~S1~c).
Apart from the absence of significant sinkholes in the regarded recharge area, leading to the fact that diffuse infiltration plays a key role for groundwater recharge, \citeA{Hacker.2003} also provided evidence for a deep karstified aquifer system with a well-connected drainage system through fractures and conduits in the Kerschbaum spring aquifer.
It is important to note that the Kerschbaum spring represents the most important source for the freshwater supply of the city and the surroundings of Waidhofen and is thus of particular interest for water resources research studies \cite{bittner2018modeling}.

\subsection{The LuKARS model}
The LuKARS model was recently proposed by \citeA{bittner2018modeling} with the aim to investigate the hydrological effects of land use changes in karst systems.
LuKARS therefore considers the dominant hydrotopes in a defined recharge area, i.\,e., areas characterized by homogeneous soil and land cover properties, as distinct spatial units.
The sum of the individual hydrotope responses to a given input signal (e.\,g., precipitation) plus the contribution of a shared linear baseflow storage is then the total spring discharge that should be modeled at a catchment's outlet.
As input data, the model itself needs a precipitation time series as well as the hydrotope soil information to run. 
If further processes affecting the effective precipitation are considered, such as interception and evapotranspiration, further input data is required.
In our case, we also take into account snow melt and accumulation, interception and evapotranspiration, for which we further need a temperature time series with a daily resolution. 
Moreover, a measured discharge time series is needed from the spring of interest to calibrate and validate the model.
In the particular case of the Kerschbaum spring, the discharge is measured with a flowmeter directly in the spring.
The discharge, precipitation, and temperature time series with a daily resolution for our model period from 2006 to 2008 were kindly provided by the water works Waidhofen a.d. Ybbs.
The input time series are shown in Fig.~S3.
The LuKARS model for the Kerschbaum spring in Waidhofen a.d. Ybbs was set up in \citeA{bittner2018modeling} and includes four spatially lumped dominant hydrotopes in the considered recharge area, shown in Fig.~S2.
Hydrotopes~1-3 have beeches as dominant tree species; however, they differ in terms of their individual soil characteristics and spatial shares.
While the first hydrotope (denoted by Hyd~1) covers 13\% of the recharge area and is characterized by shallow soils with mostly coarse-grained soil particles, hydrotope~3 (denoted by Hyd~3), in contrast, covers 27\% of the catchment and is defined by deeper and fine textured soils.
Hydrotope~2 (denoted by Hyd~2) has the largest spatial share in the Kerschbaum spring recharge area (56\%) and represents a transition hydrotope between Hyd~1 and Hyd~3 with moderate soil thicknesses and coarse to fine-textured soils.
Hydrotope~Q (denoted by Hyd~Q) characterizes the dolomite quarries, which covered about 4\% of space in the recharge area during the model period (2006-2008) in this study.

From a hydrological point of view, the areas of the dolomite quarries are drained by surface runoff and do not contribute to the Kerschbaum spring discharge.
As an obligation to avoid a possible contamination of the aquifer from the quarry areas, a protective layer consisting of fine material prevents infiltration into the groundwater system.
Thus, Hyd~Q is excluded from model calibration and will not be mentioned hereafter.
Also, \citeA{bittner2018modeling} derived the baseflow coefficient $k_{\text{b}}$ to match the relatively constant baseflow discharge of the Kerschbaum spring with its low temporal variability.
For this reason, as well as to put the focus on calibrating the hydrotope parameters, $k_{\text{b}}$ was chosen as calibrated by \citeA{bittner2018modeling}. 
More details about the LuKARS model, i.\,e., a description of the equations used in LuKARS and the relevant parameters, are provided in \ref{app:model}.
In the following, we use an index $i\in\{1,2,3\}$ to denote specifications for Hyd~$i$.

Each hydrotope is modeled as an independent bucket that has three different discharge components.
The first, representing quickflow ($Q_{\text{hyd},i}$) occurring via preferential flow paths (e.\,g., conduits), is described by a non-linear hysteresis function that is activated once a defined storage threshold ($e_{\text{max},i}$) is reached and stops after the storage value falling below a predefined minimum storage value ($e_{\text{min},i}$).
The second and third discharge components are both implemented by a linear discharge function and represent the discharge to a shared baseflow storage ($Q_{\text{is},i}$) as well as secondary spring discharge ($Q_{\text{sec},i}$), i.\,e., a discharge component that transfers water out of the catchment and does not contribute to the spring discharge.
All together, seven parameters need to be calibrated for the implementation of each single hydrotope.
These are the discharge parameter $k_{\text{hyd},i}$ and the dimensionless exponent $\alpha_i$ for $Q_{\text{hyd},i}$, the storage thresholds for the quickflow activation ($e_{\text{min},i}$) and ($e_{\text{max},i}$), parameter $k_{\text{is},i}$ as the discharge coefficient of $Q_{\text{is},i}$ and, finally, $k_{\text{sec},i}$ and $e_{\text{sec},i}$ as the discharge coefficient and the activation level for $Q_{\text{sec},i}$, respectively.
Given the different physical characteristics of all defined hydrotopes, the parameters of one hydrotope need to follow some constraints with respect to the parameters used for the implementation of other hydrotopes.
From a practical point of view, this means that a hydrotope with shallow and coarse-grained soils (e.\,g., Hyd~1) needs to have a lower storage capacity and higher discharge coefficient as compared to a hydrotope with deep and fine-textured soils (e.\,g., Hyd~3).
For the particular case of the three hydrotopes in the Kerschbaum spring recharge area, the parameter constraints are given as follows:

\begin{equation}
\label{eq:constraints}
\begin{gathered}
	k_{\text{hyd},1} \geq k_{\text{hyd},2} \geq k_{\text{hyd},3} \\
	e_{\text{min},1} \leq e_{\text{min},2} \leq e_{\text{min},3} \\
	e_{\text{max},1} \leq e_{\text{max},2} \leq e_{\text{max},3} \\
	\alpha_1 \geq \alpha_2 \geq \alpha_3 \\
	k_{\text{is},1} \geq k_{\text{is},2} \geq k_{\text{is},3} \\
	k_{\text{sec},1} \geq k_{\text{sec},2} \geq k_{\text{sec},3} \\
	e_{\text{sec},1} \leq e_{\text{sec},2} \leq e_{\text{sec},3}
\end{gathered}
\end{equation}

Although the introduced condition for the $\alpha$ values is not strictly necessary, we implemented it to further enhance the quick response of hydrotopes with a low difference between $e_{\text{min},i}$ and $e_{\text{max},i}$ and a generally low value of $e_{\text{max},i}$ during high precipitation events.
\cite{bittner2018modeling} manually calibrated the LuKARS model for the Kerschbaum spring recharge area. Based on this trial-and-error calibration, it was possible to reliably determine possible ranges of all model parameters. These are shown in Table \ref{tab:prior_interv} and will be used as prior parameter intervals for the presented study in a Bayesian setting.

\begin{table}
	\centering
	\caption{Prior intervals for physical parameters}
	\label{tab:prior_interv}
	\begin{tabular}{cccccc}
		\hline\noalign{\smallskip}
		No.  & Parameter & Lower bound & Upper bound & Unit & Description \\
		\hline\hline\noalign{\smallskip}
		$1$  & $k_{\text{hyd},1}$    & $9$       & $900$     & m$^2$d$^{-1}$          & discharge parameter for $Q_{\text{hyd},1}$  \\
		$2$  & $e_{\text{min},1}$    & $10$      & $50$      & mm                     & min. storage capacity Hyd~1  \\
		$3$  & $e_{\text{max},1}$    & $15$      & $75$      & mm                     & max. storage capacity Hyd~1  \\
		$4$  & $\alpha_{1}$          & $0.7$     & $1.6$     & --                     & quickflow exponent of Hyd~1  \\
		$5$  & $k_{\text{is},1}$     & $0.002$   & $0.2$     & m\,mm$^{-1}$d$^{-1}$   & discharge parameter for $Q_{\text{is},1}$  \\
		$6$  & $k_{\text{sec},1}$    & $0.0095$  & $0.95$    & m\,mm$^{-1}$d$^{-1}$   & discharge parameter for $Q_{\text{sec},1}$  \\
		$7$  & $e_{\text{sec},1}$    & $25$      & $70$      & mm                     & activation level for $Q_{\text{sec},1}$  \\
		\hline\noalign{\smallskip}
		$8$  & $k_{\text{hyd},2}$    & $8.5$     & $850$     & m$^2$d$^{-1}$          & discharge parameter for $Q_{\text{hyd},2}$  \\
		$9$  & $e_{\text{min},2}$    & $40$      & $80$      & mm                     & min. storage capacity Hyd~2  \\
		$10$ & $e_{\text{max},2}$    & $80$      & $160$     & mm                     & max. storage capacity Hyd~2  \\
		$11$ & $\alpha_{2}$          & $0.5$     & $1.3$     & --                     & quickflow exponent of Hyd~2  \\
		$12$ & $k_{\text{is},2}$     & $0.00055$ & $0.055$   & m\,mm$^{-1}$d$^{-1}$   & discharge parameter for $Q_{\text{is},2}$  \\
		$13$ & $k_{\text{sec},2}$    & $0.0023$  & $0.23$    & m\,mm$^{-1}$d$^{-1}$   & discharge parameter for $Q_{\text{sec},2}$  \\
		$14$ & $e_{\text{sec},2}$    & $130$     & $220$     & mm                     & activation level for $Q_{\text{sec},2}$  \\
		\hline\noalign{\smallskip}
		$15$ & $k_{\text{hyd},3}$    & $7.7$     & $770$     & m$^2$d$^{-1}$          & discharge parameter for $Q_{\text{hyd},3}$  \\
		$16$ & $e_{\text{min},3}$    & $75$      & $120$     & mm                     & min. storage capacity Hyd~3  \\
		$17$ & $e_{\text{max},3}$    & $160$     & $255$     & mm                     & max. storage capacity Hyd~3  \\
		$18$ & $\alpha_{3}$          & $0.2$     & $0.7$     & --                     & quickflow exponent of Hyd~3  \\
		$19$ & $k_{\text{is},3}$     & $0.00025$ & $0.025$   & m\,mm$^{-1}$d$^{-1}$   & discharge parameter for $Q_{\text{is},3}$  \\
		$20$ & $k_{\text{sec},3}$    & $0.0015$  & $0.15$    & m\,mm$^{-1}$d$^{-1}$   & discharge parameter for $Q_{\text{sec},3}$  \\
		$21$ & $e_{\text{sec},3}$    & $320$     & $450$     & mm                     & activation level for $Q_{\text{sec},3}$  \\
		\hline
	\end{tabular}
\end{table}

%% file: inference.tex
\section{Parameter inference}
\label{sec:inference}

In this section we present our approach for solving the Bayesian inverse problem of inferring parameter information for the LuKARS model.
Since high-dimensional parameter spaces complicate Bayesian inference, we utilize the active subspace method, a recent emerging set of tools for dimension reduction.
We focus the inference process only with a few linear combinations of parameters that are dominantly driving the update from the prior to the posterior distribution.
After the method is explained formally, but with links to the actual hydrological problem, we explain Bayesian inversion and its application in detail.

\subsection{Active subspaces}
\label{ssec:as}
The active subspace method, introduced in \citeA{constantine2014active}, \citeA{constantine2015active}, and \citeA{russi2010uncertainty}, identifies dominant directions in the domain of a multivariate, scalar-valued function $f: \mathbf{R}^n \to \mathbf{R}$.
In our context, $f$ is a data misfit function (details in Section~\ref{ssec:bi}) which quantifies the mismatch between observed and simulated spring discharge and is defined on the space of parameters to be calibrated.
In other words, we seek directions on which $f$ varies more than on other directions, on average.
Consider a function $f$ of the form $f(\mathbf{x}) = g(\mathbf{A}^\top\mathbf{x})$ for each $\mathbf{x}\in\mathbf{R}^n$, where $\mathbf{A}\in\mathbf{R}^{n\times k}$, $0<k<n$, is a rectangular matrix.
Such functions are called \textit{ridge functions}; see \citeA{pinkus2015ridge}.
Take a vector $\mathbf{v}\in\mathbf{R}^n$ from the null space of $\mathbf{A}^\top$, i.\,e., $\mathbf{A}^\top\mathbf{v}=0$, and compute
\begin{equation}
f(\mathbf{x}+\mathbf{v}) = g(\mathbf{A}^\top(\mathbf{x}+\mathbf{v})) = g(\mathbf{A}^\top\mathbf{x}) = f(\mathbf{x}).
\end{equation}
This equation shows that $f$ is constant along the null space of $\mathbf{A}^\top$ meaning that the $n$-dimensional function is actually intrinsically $k$-dimensional.
In practice, the goal is relaxed to finding approximations $g$ and $\mathbf{A}$ such that it holds that $f(\mathbf{x}) \approx g(\mathbf{A}^\top\mathbf{x})$.
For the hydrological problem of interest, it means that we try to find a few directions in the parameter space that are significantly informed by the discharge data.
We will see that some, but not all, directions change from the prior to the posterior distribution.
This fact is exploited to save a considerable amount of computational expense.

Note that this is a different approach compared to other sensitivity analysis methods like \textit{total sensitivity indices} (or \textit{Sobol indices}) \cite{sobol2001global} and \textit{Factor Priorisation} \cite{saltelli2008global}.
These methods investigate coordinate-aligned sensitivities, i.\,e., associated with a particular parameter (factor) of a hydrological model.
Active subspaces can be viewed as a generalization in the sense that we study dominant ``directions" within a parameter space, or, more precisely, we look for linear combinations of parameters that dominate the model output (here, spring discharge) on average.

In the following, we assume that $f$ is continuously differentiable and has partial derivatives that are square-inte\-grable with respect to a given probability density function $\rho$.
We study a matrix which is the $\rho$-averaged outer product of the gradient of $f$ with itself, i.\,e.,
\begin{equation}
\label{eq:C}
\mathbf{C} \coloneqq \mathbf{E}[\nabla f(\mathbf{X})\nabla f(\mathbf{X})^\top] = \int_{\mathbf{R}^n}{\nabla f(\mathbf{x})\nabla f(\mathbf{x})^\top \rho(\mathbf{x}) \; d\mathbf{x}}.
\end{equation}
In the parameter estimation problem, the weighting probability density $\rho$ is the prior density, defined for every model parameter, from the Bayesian inversion context (Section~\ref{ssec:bi}).
Note that $\mathbf{C}\in\mathbf{R}^{n\times n}$ is symmetric and positive semi-definite.
For some vector $\mathbf{v}\in\mathbf{R}^n$, compute
\begin{equation}
\mathbf{v}^\top\mathbf{C}\mathbf{v} = \mathbf{E}[(\mathbf{v}^\top \nabla f(\mathbf{X}))^2].
\end{equation}
Thus, $\mathbf{v}^\top\mathbf{C}\mathbf{v}$ displays the averaged variation of our objective function $f$ along $\mathbf{v}$.
This quantity is maximized (in the set of unit vectors) by the normalized eigenvector $\mathbf{v}_1$ of $\mathbf{C}$ corresponding to the largest eigenvalue $\lambda_1$ and gives
\begin{equation}
\mathbf{E}[(\mathbf{v}_1^\top \nabla f(\mathbf{X}))^2] = \lambda_1.
\end{equation}
For example, if $\mathbf{v}_1=\mathbf{e}_1\coloneqq(1,0,\ldots,0)^\top$, it means that $f$ is most sensitive (on average) w.r.t. changes of the first parameter.
Since small eigenvalues mean a small variation in the direction of corresponding eigenvectors, this observation suggests to compute an orthogonal eigendecomposition $\mathbf{C} = \mathbf{W}\mathbf{\Lambda}\mathbf{W}^\top$, where $\mathbf{W}=[\mathbf{w}_1,\dots,\mathbf{w}_n]$ contains the eigenvectors and $\mathbf{\Lambda} = \text{diag}(\lambda_1,\dots,\lambda_n)$ contains corresponding eigenvalues in decreasing order.
The symmetry of $\mathbf{C}$ allows us to choose $\mathbf{w}_i$ giving an orthonormal basis of $\mathbf{R}^n$.
The eigenvalues and eigenvectors can be exploited to span a lower-dimensional space along which $f$ is dominantly varying.
We can decide to split $\mathbf{W}$ after the $k$-th column and to neglect the space spanned by $\mathbf{w}_{k+1},\dots,\mathbf{w}_n$, i.\,e.,
\begin{equation}
\mathbf{W} = [\mathbf{W}_1 \;\; \mathbf{W}_2]
\end{equation}
such that $\mathbf{W}_1\in\mathbf{R}^{n\times k}$ and $\mathbf{W}_2\in\mathbf{R}^{n\times (n-k)}$.
We write
\begin{equation}
\label{eq:xyz}
\mathbf{x} = \mathbf{W}\mathbf{W}^\top\mathbf{x} = \mathbf{W}_1\mathbf{y} + \mathbf{W}_2\mathbf{z},
\end{equation}
where $\mathbf{y}\coloneqq\mathbf{W}_1^\top\mathbf{x}$ is called the \textit{active variable} and $\mathbf{z}\coloneqq\mathbf{W}_2^\top\mathbf{x}$ the \textit{inactive variable}.
The span of $\mathbf{W}_1$, i.\,e., $\mathcal{R}(\mathbf{W}_1) \coloneqq \{\mathbf{W}_1\mathbf{v} \,|\, \mathbf{v}\in\mathbf{R}^k\}$, is called the \textit{active subspace} (of $f$).
In other words, the coordinate system is transformed to a new orthogonal basis given by the eigenvectors.
The new axes corresponding to the eigenvector with the largest eigenvalue is aligned to the direction of maximum averaged variation of $f$ in the original coordinate system.

The matrix $\mathbf{C}$ is generally not available exactly in practice and must be approximated.
\citeA{constantine2014computing}, \citeA{holodnak2018probabilistic} and \citeA{lam2018multifidelity} proposed and analyzed a Monte Carlo approximation, i.\,e.,
\begin{equation}
\label{eq:C_approx}
\mathbf{C} \approx \mathbf{\tilde{C}} \coloneqq \frac{1}{N} \sum_{j=1}^{N}{\nabla f(\mathbf{X}_j) \nabla f(\mathbf{X}_j)^\top},
\end{equation}
where $\mathbf{X}_j\sim\rho$, $j=1,\dots,N>0$.
The recommended number of samples $N$ required to get a sufficiently accurate estimate of eigenvalues and eigenvectors is heuristically given by
\begin{equation}
\label{eq:N_heuristic}
N \approx \beta \, m \log(n)
\end{equation}
for a so-called \textit{sampling factor} $\beta\in[2,10]$.
The factor $m\in\mathbf{N}$ denotes the number of eigenvalues/eigenvectors to be estimated accurately.
The heuristic is motivated in \citeA{constantine2014computing} by results from random matrix theory in \citeA{tropp2012user}.
We utilize \textit{bootstrapping} \cite{constantine2014computing, efron1994introduction} to ensure that the number of gradient samples $N$ provides a sufficiently good approximation of the eigenvalues of $\tilde{\mathbf{C}}$.
Since $\mathbf{\tilde{C}}$ is only an approximation/perturbation of the exact matrix $\mathbf{C}$, eigenvalues and eigenvectors are also only available in perturbed versions, i.\,e.,
\begin{equation}
\mathbf{\tilde{C}} = \mathbf{\tilde{W}}\mathbf{\tilde{\Lambda}}\mathbf{\tilde{W}}^\top.
\end{equation}
Perturbed active and inactive variables are denoted by $\mathbf{\tilde{y}}\coloneqq\mathbf{\tilde{W}}_1^\top\mathbf{x}$ and $\mathbf{\tilde{z}}\coloneqq\mathbf{\tilde{W}}_2^\top\mathbf{x}$, respectively.

We additionally need a function $\tilde{g}$ defined on the low-dimensional (perturbed) active subspace approximating $f$ as a ridge function, i.\,e.,
\begin{equation}
f(\mathbf{x}) \approx \tilde{g}(\mathbf{\tilde{W}}_1^\top\mathbf{x})
\end{equation}
for each $\mathbf{x}\in\mathbf{R}^n$.
It is known that the best approximation in an $L^2$ sense is the conditional expectation conditioned on the active variable $\mathbf{\tilde{y}}$, i.\,e.,
\begin{equation}
\tilde{g}(\mathbf{\tilde{y}}) = \int_{\mathbf{R}^{n-k}}{f(\mathbf{\tilde{W}}_1\mathbf{\tilde{y}} + \mathbf{\tilde{W}}_2\mathbf{\tilde{z}}) \, \rho_{\mathbf{\tilde{Z}}|\mathbf{\tilde{Y}}}(\mathbf{\tilde{z}}|\mathbf{\tilde{y}}) \; d\mathbf{\tilde{z}}}.
\end{equation}
The conditional probability density function $\rho_{\mathbf{\tilde{Z}}|\mathbf{\tilde{Y}}}$ is defined in the usual way, see e.\,g., \citeA[Section 20 and 33]{billingsley1995probability}.

In Section~\ref{sec:results}, we make use of a cheap response surface to $\tilde{g}$ gained by a polynomial regression approach since evaluating $\tilde{g}$, or even a Monte Carlo approximation of it, can get costly due to additional evaluations of $f$ required.
This surrogate is constructed according to instructions described in Algorithm~\ref{alg:surrogate_g}.
There are several examples in the literature that show that a polynomial approximation can be useful in the context of active subspaces, e.\,g., \cite{cortesi2017forwardbackward, teixeiraparente2019efficient}.
The accuracy of a regression fit is measured by the $r^2$ value, or \textit{coefficient of determination} (see, for example, \citeA{glantz1990primer}).

\begin{algorithm}
	\caption{Response surface construction}
	\label{alg:surrogate_g}
	\begin{flushleft}
		Assume $M>0$ samples $\mathbf{x}_i$, $i=1,\dots,M,$ according to $\rho$ and corresponding function values $f_i$, $i=1,\dots,M,$ are given.
		\begin{enumerate}
			\item Compute samples $\mathbf{\tilde{y}}_i$ in the active subspace by
			\begin{equation}
			\mathbf{\tilde{y}}_i = \mathbf{\tilde{W}}_1^\top \mathbf{x}_i, \quad i=1,\dots,M.
			\end{equation}
			\item Find a regression surface $\tilde{G}$ for pairs $(\mathbf{\tilde{y}}_i,f_i)$ such that
			\begin{equation}
			\tilde{G}(\mathbf{\tilde{y}}_i) \approx f_i, \quad i=1,\ldots,M.
			\end{equation}
			\item Get a low-dimensional approximation of $f$ at $\mathbf{x}$ by computing
			\begin{equation}
			f(\mathbf{x}) \approx \tilde{G}(\mathbf{\tilde{W}}_1^\top \mathbf{x}).
			\end{equation}
		\end{enumerate}
	\end{flushleft}
\end{algorithm}

\subsection{Global sensitivity analysis with active subspace}
\label{ssec:sens_analy}
\citeA{constantine2017global} show that it is possible to get global sensitivity values from the active subspace that are comparable, in practical situations, with more familiar metrics like \textit{variance-based sensitivities}, also known as \textit{total sensitivity indices} or \textit{Sobol indices} \cite{saltelli2008global,sobol2001global}.

Since the expensive computations for building the matrix $\mathbf{\tilde{C}}$ are already done, no further huge computational costs are needed.
By ``global'' we mean that the sensitivities, assigned to each parameter individually, are averaged quantities.
In particular, the matrix $\mathbf{C}$ from Eq.~\eqref{eq:C}, which will be exploited to compute global sensitivities, is constructed with gradients of the function of interest $f$ at different locations weighted with a given probability density $\rho$.
For our application of the LuKARS model, the function~$f$ and the density~$\rho$ are taken to be data misfit function and the prior density of the model parameters from the Bayesian context described in Section~\ref{ssec:bi}.

The vector of sensitivities $\bm{\mathbf{s}}\in\mathbf{R}^n$, in which the $i$-th component displays the (global) sensitivity of $f$ w.r.t. parameter $\mathbf{x}_i$, is in \citeA{constantine2017global} computed via
\begin{equation}
s_i \coloneqq s_i(m) \coloneqq \sum_{j=1}^{m}{\lambda_j w_{i,j}^2}, \quad i=1,\dots,m, \; 0 < m \leq n.
\end{equation}
Here, we will set $m=n$.
Thus, we can write more compactly
\begin{equation}
\mathbf{s}(n) = (\mathbf{W}\circ\mathbf{W}) \bm{\lambda},
\end{equation}
where $\bm{\lambda}=(\lambda_1,\dots,\lambda_n)^\top\in\mathbf{R}^n$ is the vector of eigenvalues and $\circ$ denotes elementwise multiplication.

Similarly to the estimated quantities in previous sections, we will only have an estimate $\mathbf{\tilde{s}}$ available due to the finite approximation of $\mathbf{C}$.
In general, it is hard to give strict bounds for the number of samples $N$ required to get a sufficiently accurate approximation to $\mathbf{s}$.
Hence, we use as many samples as were shown to be sufficient in \cite{constantine2017global}.

\subsection{Bayesian inversion}
\label{ssec:bi}
The aim of Bayesian inversion is to approximate a \textit{posterior} probability distribution on the space of parameters $\mathbf{x}\in\mathbf{R}^n$, $n\in\mathbf{N}$, that incorporates uncertainty in the estimated parameters due to noise in the measured discharge data.
\citeA{stuart2010inverse} gives a rigorous mathematical framework for Bayesian inverse problems, even in infinite-dimensional parameter spaces.
The starting point in Bayesian inversion is a \textit{prior} probability distribution $\rho_{\text{prior}}$ that serves as a first guess on the distribution of the model parameters \textit{without} any incorporation of measured hydrological data.
The prior also serves to regularize the inverse problem.
This choice is often driven by intuition or expert knowledge.
Mathematically speaking, we seek a distribution on $\mathbf{x}$ conditioned on the observation of specific measured data.
This leads directly to the well-known Bayes' theorem.

Data $\mathbf{d}\in\mathbf{R}^{n_\mathbf{d}}$, $n_\mathbf{d}\in\mathbf{N}$, are here modeled as
\begin{equation}
\mathbf{d} = \mathcal{G}(\mathbf{x}) + \bm{\eta},
\end{equation}
where $\bm{\eta}\sim\mathcal{N}(0,\Gamma)$ is additive Gaussian noise, modeling measurement errors, with mean zero and covariance matrix $\Gamma \in \mathbf{R}^{n_\mathbf{d} \times n_\mathbf{d}}$ and $\mathcal{G}:\mathbf{R}^n \to \mathbf{R}^{n_\mathbf{d}}$ is called the \textit{parameter-to-observation map}.
This map is composed of a \textit{forward operator} $G:\mathbf{R}^n \to V$, displaying, e.\,g., the solution to a partial differential equation (PDE), and an \textit{observation operator} $\mathcal{O}:V \to \mathbf{R}^{n_\mathbf{d}}$, being, e.\,g., a linear functional on the PDE solution space $V$.
For the LuKARS model, $\mathcal{G}$ is the mapping from the calibration parameters $\mathbf{x}$ (related to parameters in Table~\ref{tab:prior_interv} and described in Section~\ref{ssec:calib_param}) to the discharge values.
By Bayes' theorem, we can define the posterior density as
\begin{equation}
\rho_{\text{post}}(\mathbf{x}) \coloneqq \rho_{\text{post}}(\mathbf{x}|\mathbf{d}) = \frac{\rho_{\text{like}}(\mathbf{x}) \, \rho_{\text{prior}}(\mathbf{x})}{Z},
\end{equation}
where $Z \coloneqq \int_{\mathbf{R}^n}{\rho_{\text{like}}(\mathbf{x}') \, \rho_{\text{prior}}(\mathbf{x}')} \, d\mathbf{x}'$ is a normalizing constant to get a proper probability density with unit mass.
The \textit{likelihood} $\rho_{\text{like}}$ denotes the probability that a parameter $\mathbf{x}$ is explaining the discharge data $\mathbf{d}$ corrupted by noise.
In this context, i.\,e., assuming additive Gaussian noise, the likelihood is given by
\begin{equation}
	\label{eq:likeli}
	\rho_{\text{like}}(\mathbf{x}) \propto \exp(-f_{\mathbf{d}}(\mathbf{x}))
\end{equation}
with the \textit{data misfit function} $f_\mathbf{d}(\mathbf{x}) \coloneqq \frac{1}{2} \lVert \mathbf{d}-\mathcal{G}(\mathbf{x}) \rVert_\Gamma^2$ and $\lVert \cdot \rVert_\Gamma \coloneqq \lVert \Gamma^{-1/2}\cdot \rVert_2$.
Note that the data misfit function is not a typical squared error function, but involves weights by the noise covariance matrix $\Gamma$.

The posterior density is often intractable since its evaluation requires the solution of a potentially computationally intense problem hidden in the forward operator~$G$.
The situation becomes even worse if the inverse problem is stated in a high-dimensional parameter space.
Whereas a single run of the LuKARS model is sufficiently cheap in our case study, the issue is the high-dimensionality of the problem.
A common way to approximate an expensive posterior distribution is to construct samples distributed according to the posterior.
However, many sampling techniques suffer from the curse of dimensionality.
Well-known sampling approaches comprise, e.\,g., Markov chain Monte Carlo (MCMC), \textit{Sequential Monte Carlo} (SMC), \textit{Importance Sampling}, and combinations of them.

In this work, we use a \textit{Metropolis-Hastings algorithm} from \citeA{hastings1970monte} which belongs to the class of MCMC methods.
The algorithm constructs a discrete Markov chain whose components are taken as samples and are stationarily distributed according to the desired distribution which is the posterior here.
The samples are naturally correlated which is a drawback compared to other sampling techniques that produce independent samples.
However, advantages of this algorithm are the absence of restricting assumptions and the fact that it does not suffer from the curse of dimensionality as badly as other samplers.
Nevertheless, MCMC methods can have deteriorating behavior in higher dimensions because the number of steps needed to get a sufficiently small correlation between two samples can be rather large.
Since the forward operator $G$ is evaluated in every step in the Metropolis-Hastings algorithm, the standard usage of the algorithm can get computationally expensive, especially if $G$ is costly.

In this manuscript, we run a standard Metropolis-Hastings algorithm in low dimensions.
For finding low-dimensional structure in our problem, we apply the active subspace method, described in the previous subsection, which allows to find dominant directions in a parameter space that drive the update from the prior to the posterior distribution in Bayesian inverse problems.
Additionally, it provides a cheap surrogate of the data misfit function in the low-dimensional space (see Section~\ref{ssec:as}).

\subsection{MCMC in the active subspace}
\label{ssec:mcmc_as}
For Bayesian inversion, the function of interest that we aim to approximate with a low-dimensional approximation is the data misfit function $f_\mathbf{d}$ from Eq.~\eqref{eq:likeli}, i.\,e.,
\begin{equation}
	f_\mathbf{d}(\mathbf{x}) \coloneqq \frac{1}{2}\lVert \mathbf{d}-\mathcal{G}(\mathbf{x}) \rVert_\Gamma^2.
\end{equation}
The gradient of $f_\mathbf{d}$ needed for the computation of $\mathbf{\tilde{C}}$ is
\begin{equation}
	\label{eq:grad_f}
	\nabla f_\mathbf{d}(\mathbf{x}) = \nabla\mathcal{G}(\mathbf{x})^\top\Gamma^{-1}(\mathcal{G}(\mathbf{x}) - \mathbf{d}),
\end{equation}
where $\nabla\mathcal{G}$ denotes the Jacobian matrix of the parameter-to-observation operator $\mathcal{G}$ which models the relationship between model parameters and discharge values in the LuKARS model.

Not only the perturbed active subspace, but also a cheap surrogate $\tilde{G}_\mathbf{d}$ for $\tilde{g}_\mathbf{d}$, given by
\begin{equation}
\tilde{g}_\mathbf{d}(\mathbf{\tilde{y}}) = \int_{\mathbf{R}^{n-k}}{f_\mathbf{d}(\mathbf{\tilde{W}}_1\mathbf{\tilde{y}}+\mathbf{\tilde{W}}_2\mathbf{\tilde{z}}) \, \rho_{\mathbf{\tilde{Z}}|\mathbf{\tilde{Y}}}(\mathbf{\tilde{z}} | \mathbf{\tilde{y}}) \; d\mathbf{\tilde{z}}},
\end{equation}
can be exploited for an accelerated MCMC algorithm in lower dimensions producing posterior samples for a Bayesian inverse problem as shown in \citeA{constantine2016accelerating}.
The construction of $\tilde{G}_\mathbf{d}$ as a polynomial approximation is described in Algorithm~\ref{alg:surrogate_g}.
Note that there are no additional full model evaluations necessary for this construction.
The full model evaluations that we get as a byproduct of the gradient calculations (see Eq.~\eqref{eq:grad_f}) can be reused.
By acceleration, we mean that the mixing behavior of the resulting Markov chains constructed by a standard Metropolis-Hastings algorithm can improve in lower dimensions.
As a consequence, the computational effort to produce a certain number of posterior samples is reduced.

In a first step, we compute samples of the posterior distribution defined on the low-dimensional subspace, called \textit{active posterior samples}.
In order to evaluate the (approximate) posterior density $\tilde{\rho}_{\text{post},\mathbf{\tilde{Y}}}$ in the active subspace given by
\begin{equation}
	\rho_{\text{post},\mathbf{\tilde{Y}}}(\mathbf{\tilde{y}}) \approx \tilde{\rho}_{\text{post},\mathbf{\tilde{Y}}}(\mathbf{\tilde{y}}) \propto \exp(-\tilde{G}_\mathbf{d}(\mathbf{\tilde{y}})) \, \rho_{\text{prior},\mathbf{\tilde{Y}}}(\mathbf{\tilde{y}}),
\end{equation}
where $\tilde{G}_\mathbf{d}$ is the response surface approximating $\tilde{g}_\mathbf{d}$, we need an approximation to $\rho_{\text{prior},\mathbf{\tilde{Y}}}$ denoting the marginal prior density on the perturbed active variable.
The marginal prior density is in general not analytically available and has to be estimated, e.\,g., with kernel density estimation (KDE).
Note that for the Metropolis-Hastings algorithm it is only important to know the density up to a constant.
The algorithmic details are given in Algorithm~\ref{alg:mcmc_as}.

\begin{algorithm}
	\caption{MCMC in the active subspace}
	\label{alg:mcmc_as}
	\begin{flushleft}
		Assume a state $\mathbf{\tilde{y}}_i$ is given at step $i$.
		Let $\rho(\cdot\,|\,\mathbf{\tilde{y}}_i)$ be a symmetric proposal density function and denote the surrogate on $\tilde{g}_\mathbf{d}$ with $\tilde{G}_\mathbf{d}$.
		Furthermore, suppose a density $\hat{\rho}$ estimating $\rho_{\text{prior},\mathbf{\tilde{Y}}}$ is given.
		Then, one step of the algorithm is:
		\begin{enumerate}
			\item Propose a candidate $\mathbf{y}'\sim\rho(\cdot\,|\,\mathbf{\tilde{y}}_i)$.
			\item Compute the acceptance probability with
				\begin{equation}
					\gamma(\mathbf{y}',\mathbf{\tilde{y}}_i) \coloneqq \min\left\lbrace 1, \frac{\exp(-\tilde{G}_\mathbf{d}(\mathbf{y}')) \, \hat{\rho}(\mathbf{y}')}{\exp(-\tilde{G}_\mathbf{d}(\mathbf{\tilde{y}}_i)) \, \hat{\rho}(\mathbf{\tilde{y}}_i)} \right\rbrace.
				\end{equation}
			\item Draw a uniform sample $u\sim\mathcal{U}(0,1]$.
			\item Accept/reject $\mathbf{y}'$ according to $u \leq \gamma(\mathbf{y}',\mathbf{\tilde{y}}_i)$.
		\end{enumerate}
	\end{flushleft}
\end{algorithm}

The active posterior samples are naturally correlated.
Nevertheless, the autocorrelation time is much lower compared to higher dimensional Markov chains.
The so-called \textit{effective sample size} (ESS) displaying estimation quality of a sequence of samples can be computed by a formula from \citeA{brooks2011handbook},
\begin{equation}
	N_{\mathbf{\tilde{y}}^{(\ell)}, \text{ESS}} = \frac{N_\mathbf{\tilde{y}}}{1+2\sum_{j=1}^{J_{\text{max}}}{r^{(\ell)}_j}}.
\end{equation}
The $\ell$-th component of a sample $\mathbf{\tilde{y}}$ is denoted by $\mathbf{\tilde{y}}^{(\ell)}$ and the number of samples available by $N_{\mathbf{\tilde{y}}}$.
The expression $r^{(\ell)}_j$ describes the autocorrelation between the $\ell$-th component of samples $\mathbf{\tilde{y}}$ with lag $j$.
The maximum lag regarded is given by $J_{\text{max}}$.
We determine the final effective sample size with
\begin{equation}
	N_\text{ESS} = \min_{\ell=1,\dots,k}{N_{\mathbf{\tilde{y}}^{(\ell)},\text{ESS}}}.
\end{equation}
After completing Algorithm~\ref{alg:mcmc_as}, the entire set of active posterior samples $\tilde{\mathbf{y}}$ is reduced to a set of size $N_\text{ESS}$ by taking every $p$-th sample in the chain (where $p$ is chosen such that we get a set of size $N_\text{ESS}$).
The auto-correlation of the chosen samples is thus reduced.
This technique is called \textit{thinning} (see, e.\,g., \cite{link2012thinning}).
The adequacy of the size of samples depends on their application in general.
However, it is possible to check it, for example by bootstrapping or, if the surrogate is cheap enough, by construction and comparison of multiple Markov chains having different lengths.

Our final goal is to construct samples of the posterior distribution in the original $n$-dimensional space.
Eq.~\eqref{eq:xyz} suggests to sample $\mathbf{\tilde{z}}$-components for each $\mathbf{\tilde{y}}$ from the reduced set of samples gained from Algorithm~\ref{alg:mcmc_as}.
Since it is generally not trivial to sample from the conditional distribution of $\mathbf{\tilde{z}}$ given $\mathbf{\tilde{y}}$, $\rho_{\mathbf{\tilde{Z}}|\mathbf{\tilde{Y}}}(\mathbf{\tilde{z}}|\mathbf{\tilde{y}})$, we have to run another Metropolis-Hastings algorithm.
Note that for this sampling, we only need to sample from $\rho_{\text{prior}}(\mathbf{\tilde{W}}_1\mathbf{\tilde{y}} + \mathbf{\tilde{W}}_2\mathbf{\tilde{z}})$ because $\rho_{\mathbf{\tilde{Z}}|\mathbf{\tilde{Y}}}$ is proportional to that distribution.
We again compute (nearly uncorrelated) samples of $\mathbf{\tilde{z}}$ given a particular sample $\mathbf{\tilde{y}}$ to finally translate them to posterior samples $\mathbf{x} = \mathbf{\tilde{W}}_1\mathbf{\tilde{y}} + \mathbf{\tilde{W}}_2\mathbf{\tilde{z}}$ in the original full-dimen\-sional parameter space.

\begin{figure}
	\centering
	\includegraphics[width=\textwidth]{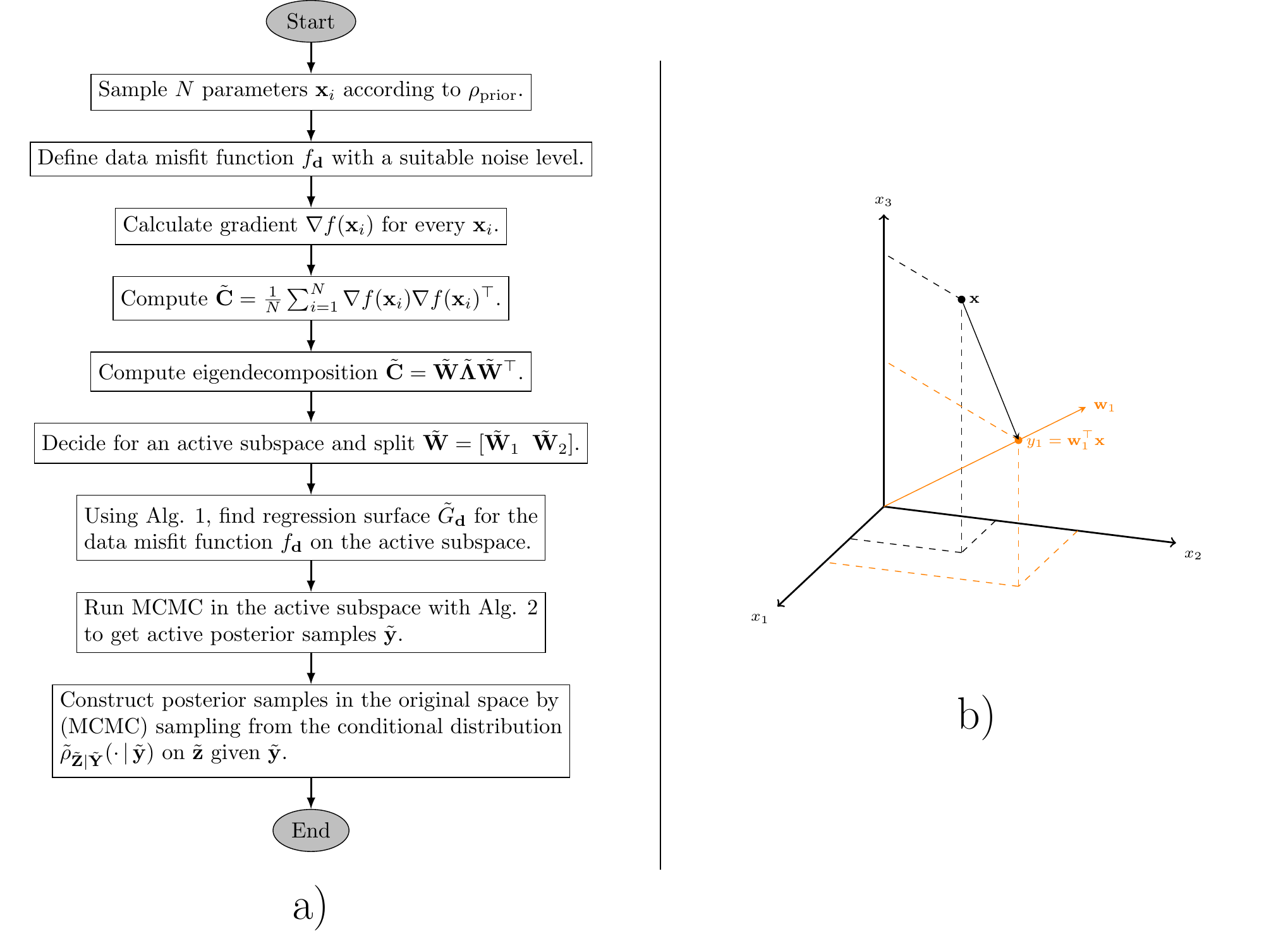}
	\caption{a) Flowchart displaying the main steps in the present Bayesian inference process with active subspaces.
	b) Visualization of a one-dimensional active subspace in a three-dimensional parameter space.
	The parameter $\mathbf{x}\in\mathbf{R}^3$ is projected on the active subspace spanned by $\mathbf{w}_1$ giving the active variable $y_1=\mathbf{w}_1^\top\mathbf{x}$.}
	\label{fig:flowchart_act_subsp_3D}
\end{figure}

A summarizing flowchart displaying the described steps of our approach is depicted in Fig.~\ref{fig:flowchart_act_subsp_3D}~a).
We emphasize that full model evaluations are only necessary for the construction of the active subspace.
The surrogate can be constructed with model evaluations $\mathcal{G}(\mathbf{x}_i)$ that we get as a byproduct from the calculation of gradients, see Eq.~\eqref{eq:grad_f}.
However, if it is required to perform a check for the potential issue of overfitting in the regression fit, additional model evaluations are necessary.
For example, we computed another 20,000 model simulations to prevent overfitting, although the original 1,000 evaluations would have been enough (see Section~\ref{sec:results}).
Eventually, Fig.~\ref{fig:flowchart_act_subsp_3D}~b) visualizes the fact that the active subspace method is looking for dominant \textit{directions} in a high-dimensional parameter spaces instead of studying sensitivities of coordinate-aligned, individual parameters.

\subsection{Parameter setting}
\label{ssec:calib_param}
Before we describe the results of the parameter study in the next section, we discuss the calibrated parameters and their notation.
As mentioned in Section~\ref{sec:model_data}, there are three hydrotopes with 7 variable parameters each.
These parameters are called \textit{physical parameters} in the following.
All of the 7 parameters have the same physical meaning for each hydrotope.

There are two reasons that lead to the introduction of artificial parameters which we call \textit{calibration parameters} in the following.
One reason is that the $k_*$ values are calibrated on a log scale.
Therefore, we define
\begin{equation}
k^\text{log}_* = \log(k_*)
\end{equation}
for each $k_* \in \{k_{\text{hyd},i}, k_{\text{is},i}, k_{\text{sec},i}\}$, $i=1,2,3$.
The second reason is the dependence of the physical parameters which needs to be circumvented since the application of active subspaces in Bayesian inverse problems prefers independently distributed and normalized parameters.
There exist two types of dependencies, namely
\begin{enumerate}
	\item cross-hydrotope dependencies caused by Eq.~\eqref{eq:constraints} \label{it:cross_dep}
	\item a dependence between parameters $e_{\text{min},i}$ and $e_{\text{max},i}$, since $e_{\text{min},i} \leq e_{\text{max},i}$.
\end{enumerate}
The first point concerns only parameters in hydrotope 2 and 3.
Hence, we write
\begin{equation}
p_i = p_{i,\text{lb}} + \triangle p_{(i-1,i)} (\min\{p_{i,\text{ub}},p_{i-1}\} - p_{i,\text{lb}})
\end{equation}
or
\begin{equation}
p_i = \max\{p_{i-1},p_{i,\text{lb}}\} + \triangle p_{(i-1,i)} (p_{i,\text{ub}} - \max\{p_{i-1},p_{i,\text{lb}}\}), \quad i=2,3,
\end{equation}
depending on whether the physical parameter $p_i \in \{k^\text{log}_{\text{hyd},i}, e_{\text{min},i}, \alpha_i, k^\text{log}_{\text{is},i}, k^\text{log}_{\text{sec},i}, e_{\text{sec},i}\}$ follows a decreasing or, respectively, increasing behavior (see Eq.~\eqref{eq:constraints}).
The fixed values $p_{i,\text{lb}}$ and $p_{i,\text{ub}}$ denote the lower and upper bound of respective physical parameters given in Table~\ref{tab:prior_interv}.
The parameters $\triangle p_i \in [0,1]$ are the newly introduced calibration parameters.
They are independent of other calibration parameters.
For the second point, we replace (for the purpose of calibration) the parameters $e_{\text{max},i}$ by $\triangle e_i$ and set
\begin{equation}
\label{eq:e_minmax}
e_{\text{max},i} = e_{\text{min},i} + \triangle e_i, \quad i=1,2,3.
\end{equation}
The parameter $\triangle e_i$ is now independent of $e_{\text{min},i}$.
Minimum and maximum values for $\triangle e_i$ are computed with respective intervals from Table~\ref{tab:prior_interv}, i.\,e., we have
\begin{equation}
e_{\text{max},i,\text{lb}} - e_{\text{min},i,\text{lb}} \leq \triangle e_i \leq e_{\text{max},i,\text{ub}} - e_{\text{min},i,\text{ub}},
\end{equation}
which is valid since it holds that $e_{\text{max},i,\text{lb}} - e_{\text{min},i,\text{lb}} \leq e_{\text{max},i,\text{ub}} - e_{\text{min},i,\text{ub}}$ for every $i=1,2,3$.

Finally, all calibration parameters need to be normalized, i.\,e., they are mapped from their corresponding interval to $[-1,1]$.
Normalized parameters are denoted with a bar.
Summarizing, the vector of all normalized independent calibration parameters is
\begin{align}
\mathbf{x} = (& \bar{k}_{\text{hyd},1}^{\text{log}}, \bar{e}_{\text{min},1}, \triangle \bar{e}_1, \bar{\alpha}_{1}, \bar{k}_{\text{is},1}^{\text{log}}, \bar{k}_{\text{sec},1}^{\text{log}}, \bar{e}_{\text{sec},1}, \notag \\
& \triangle \bar{k}_{\text{hyd},(1,2)}^{\text{log}}, \triangle \bar{e}_{\text{min},(1,2)}, \triangle \bar{e}_2, \triangle \bar{\alpha}_{(1,2)}, \triangle \bar{k}_{\text{is},(1,2)}^{\text{log}}, \triangle \bar{k}_{\text{sec},(1,2)}^{\text{log}}, \triangle \bar{e}_{\text{sec},(1,2)}, \label{eq:param_calib} \\
& \triangle \bar{k}_{\text{hyd},(2,3)}^{\text{log}}, \triangle \bar{e}_{\text{min},(2,3)}, \triangle \bar{e}_3, \triangle \bar{\alpha}_{(2,3)}, \triangle \bar{k}_{\text{is},(2,3)}^{\text{log}}, \triangle \bar{k}_{\text{sec},(2,3)}^{\text{log}}, \triangle \bar{e}_{\text{sec},(2,3)})^\top \in \mathbf{R}^{21}. \notag
\end{align}

%% file: results.tex
\section{Results}
\label{sec:results}
In the following, we assume a normally distributed measurement error (noise) at a level of $5\%$ for the measured spring discharge as it was kindly provided by the water works owner Waidhofen a.d. Ybbs for the applied flowmeter. This translates to $\Gamma_{ij} = (0.05 \times d_i)^2\delta_{ij}$, $i,j=1,\dots,n_\mathbf{d}$, where $\Gamma$ is the covariance matrix.
Additionally, we assume a uniform distribution on the calibration parameters from Eq.~\eqref{eq:param_calib}, i.\,e., $\mathbf{x}\sim\mathcal{U}[-1,1]^{21}$.
The prior intervals for the physical parameters are given in Table~\ref{tab:prior_interv}.
Note that the prior distribution on the physical parameters is not uniform due to the transformation described in Section~\ref{ssec:calib_param}.

For the computation of $\mathbf{\tilde{C}}$ from Eq.~\eqref{eq:C_approx}, we use $N=1,000$ gradient samples of $\nabla f_\mathbf{d}$, although only about 250 samples would be necessary to estimate the first $m=8$ eigenvectors sufficiently accurately according to Eq.~\eqref{eq:N_heuristic} with a pessimistic sampling factor $\beta = 10$.
The reason for choosing this rather large number is to make sure that the global sensitivity values, for which such heuristics do not exist, are also estimated accurately.
The gradient was approximated by central finite differences.
Using seven cores of type Intel(R) Xeon(R) E5 at 3\,GHz each, the required $1,000 \times (21\times2+1) = 43,000$ forward runs need about 4.3 hours since it required 2.5 seconds for a single run of the model.

\begin{figure}
	\centering
	\includegraphics[width=0.7\textwidth]{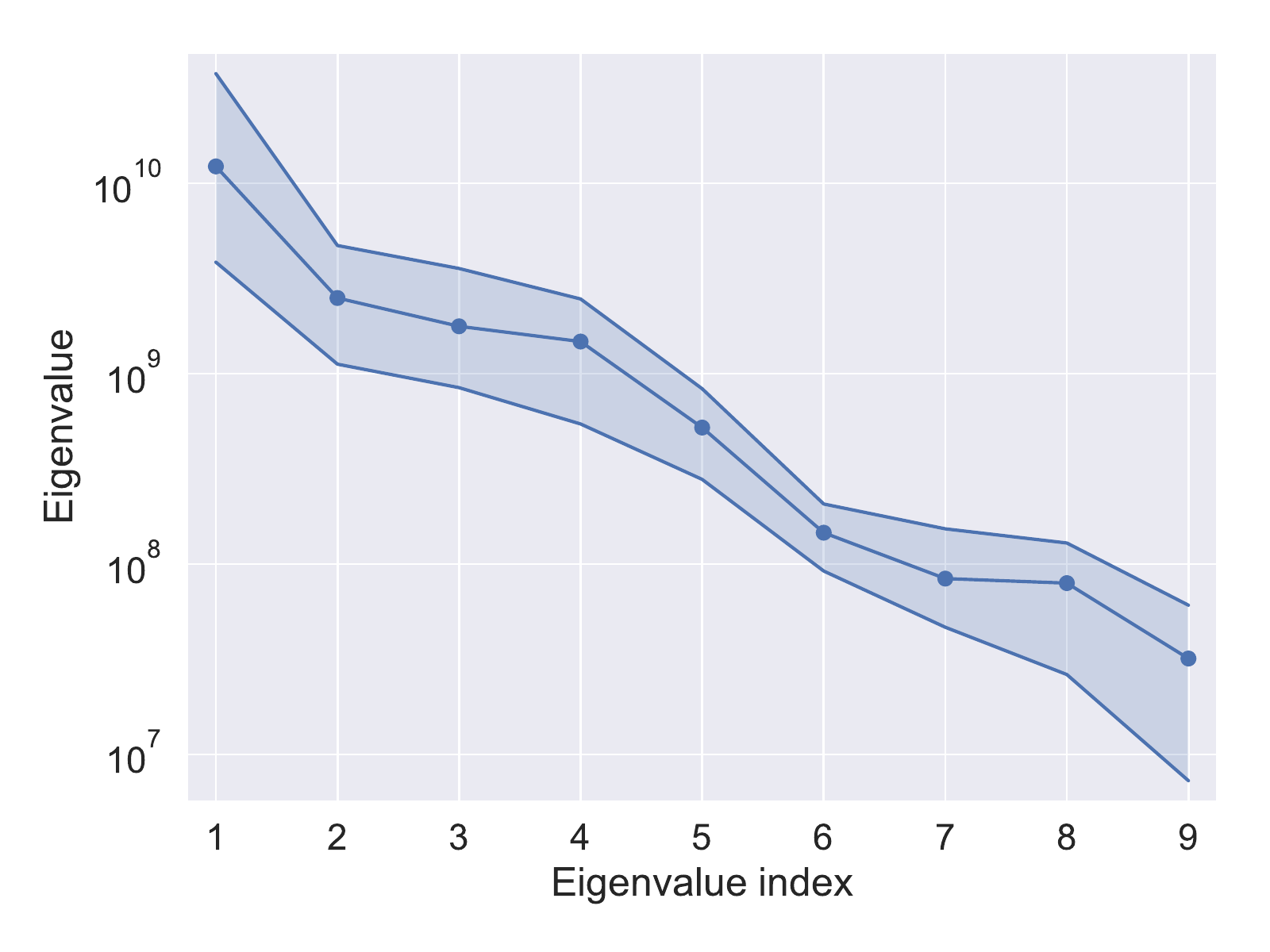}
	\caption{Spectrum of the matrix $\mathbf{\tilde{C}}$ for the data misfit function $f_\mathbf{d}$ with a 5\% noise level.
		The light blue area around the eigenvalues indicates variability of the eigenvalues caused by the random nature of the approximation $\tilde{\mathbf{C}}$.}
	\label{fig:eigvals_4d_5}
\end{figure}

\begin{figure}
	\centering
	\includegraphics[width=\textwidth]{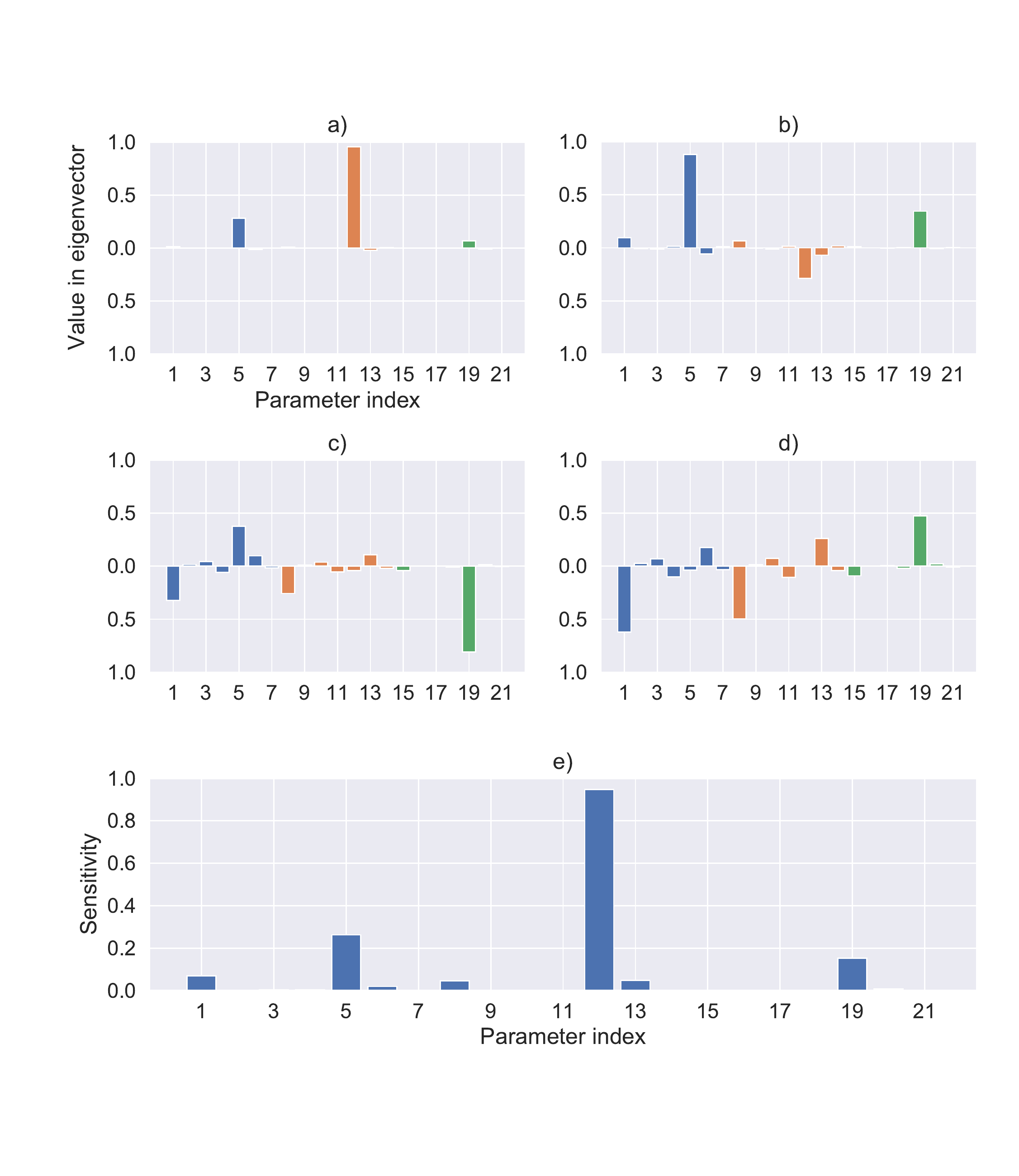}
	\caption{The subfigures show sensitivities with respect to the calibration parameters described in Section~\ref{ssec:calib_param}.
		First two rows: First four eigenvectors of the matrix $\mathbf{\tilde{C}}$ ($\tilde{\mathbf{w}}_i$, $i=1,2,3,4$) for the data misfit function $f_\mathbf{d}$ with a 5\% noise level.
		Along these directions, the data misfit function changes much more, on average, than along other directions (eigenvectors).
		The colors distinguish the three hydrotopes (blue: Hyd~1, orange: Hyd~2, green: Hyd~3).
	Last row: Global sensitivity values of the data misfit function $f_\mathbf{d}$ with a 5\% noise level.
	These values are computed using the eigenvalues and eigenvectors of $\tilde{\mathbf{C}}$ described in Section~\ref{ssec:sens_analy}.
	The ratio of maximum and minimum sensitivities is $3.6\times10^6$.}
	\label{fig:eigvecs_4d_5_scores_5}
\end{figure}

The resulting eigenvalues and first four eigenvectors are plotted in Fig.~\ref{fig:eigvals_4d_5} and Fig.~\ref{fig:eigvecs_4d_5_scores_5}~a)-d), respectively. 
Fig.~\ref{fig:eigvals_4d_5} shows the spectral decay on a logarithmic scale.
The light blue area around the eigenvalues displays minimum and maximum eigenvalues gained from bootstrapping and indicates the variability of eigenvalues due to the random nature of the approximation $\tilde{\mathbf{C}}$ (see Eq.~\eqref{eq:C_approx}).
Gaps after the first and fourth eigenvalue suggest the existence of one- and four-dimensional active subspaces.
Fig.~\ref{fig:eigvecs_4d_5_scores_5}~a)-d) shows eigenvectors with components colored according to the hydrotope they are supposed to model.
It shows that parameters 5, 12, and 19, having large contributions in the first three eigenvectors, take a dominant role.
All of these parameters involve the $k_{\text{is}}$ value for each hydrotope.
Moreover, we can observe a ranking between the corresponding hydrotopes of these parameters, with decreasing order of the $k_{\text{is}}$ values from Hyd~2, Hyd~1 to Hyd~3 in the first eigenvector.
A different pattern is observed for the contributions of parameters related to $k_{\text{hyd}}$, represented by parameters 1, 8, and 15 for Hyd~1, Hyd~2, and Hyd~3, respectively.
These parameters appear in eigenvectors 2 to 4 and show a different ranking as compared to the $k_{\text{is}}$ parameters, with a decreasing contribution from Hyd~1, Hyd~2 to Hyd~3.
It is important to highlight that parameter 15 only shows a small contribution in eigenvector 3 and 4.
A third important parameter group is related to $k_\text{sec}$, here represented by parameters 6, 13, and 20.
They appear in eigenvectors 2 to 4 and have comparable contributions in Hyd~1 and Hyd~2 (parameters 6 and 13) but only a minor contribution in Hyd~3 (see eigenvector 4).
Interestingly, for hydrotope 1 and 2 the same parameters show up with a similar shape in eigenvector $\mathbf{\tilde{w}}_4$.

It is important to emphasize that the resulting eigenvectors for the Bayesian data misfit can look different to ones gained from using the \textit{Nash-Sutcliffe model efficiency coefficient} (NSE), which is a more common misfit function in hydrology, introduced in \citeA{nash1970river}.
This can be explained by the fact that the NSE is focusing on high-flow conditions as it is a (non-weighting) squared error function \cite{gupta2009decomposition}, whereas the Bayesian data misfit function is a weighting/relative squared error function (weights are given by entries in the noise covariance matrix $\Gamma$) and, thus, does not favor low- or high-flow conditions.

With the eigenvalue/-vector plot, we have already gained some insight in the parameter sensitivities.
Fig.~\ref{fig:eigvecs_4d_5_scores_5} e) shows the global sensitivities of the data misfit function $f_\mathbf{d}$ normalized to $[0,1]$.
The most sensitive parameter is $\triangle k_{\text{is},(1,2)}$, but also $\bar{k}_{\text{is},1}$ and $\triangle k_{\text{is},(2,3)}$ have their contributions since they show up in $\mathbf{\tilde{w}}_2$ and $\mathbf{\tilde{w}}_3$, respectively, with non-negligible corresponding eigenvalues.
At the same time, parameters 6, 13, and 20, involving $k_\text{sec}$ values in the hydrotopes, show sensitive in the first eigenvectors.
Parameters displaying $k_\text{hyd}$ values have small contributions but only in hydrotope 1 and 2.
All other parameters do not show much sensitivity since their components contribute only to eigenvectors having eigenvalues that are orders of magnitudes smaller than the first four.
As a consequence, we expect that the more sensitive parameters change their distribution (and also joint distributions) from the prior to the posterior during Bayesian inference.

We decide for a 4D subspace and compute a 4th order polynomial to get the response surface $\tilde{G}_\mathbf{d}$ of $\tilde{g}_\mathbf{d}$ by Algorithm~\ref{alg:surrogate_g}.
Since polynomials of 4th order already have $70$ degrees of freedom in four dimensions, we compute another 20,000 samples of $f_\mathbf{d}(\mathbf{x})$, with $\mathbf{x}$ following the prior, to preclude overfitting.
Nevertheless, the 1,000 samples from the computation of $\mathbf{\tilde{C}}$ would have been enough to get the same $r^2$ score which is $\approx 0.77$.
This score, also called coefficient of determination, is a statistical measure for the goodness of a fit and reflects the percent of variance explained.
If predictions of a regression match perfectly well with the data points, the $r^2$ score becomes 1.
In contrast, it can become less than zero, if the predicted values are worse than choosing the constant mean value of the data.
In this regard, our $r^2$ score indicates that our surrogate is a sufficiently well behaved fit.

\begin{figure}
	\centering
	\includegraphics[width=0.7\textwidth]{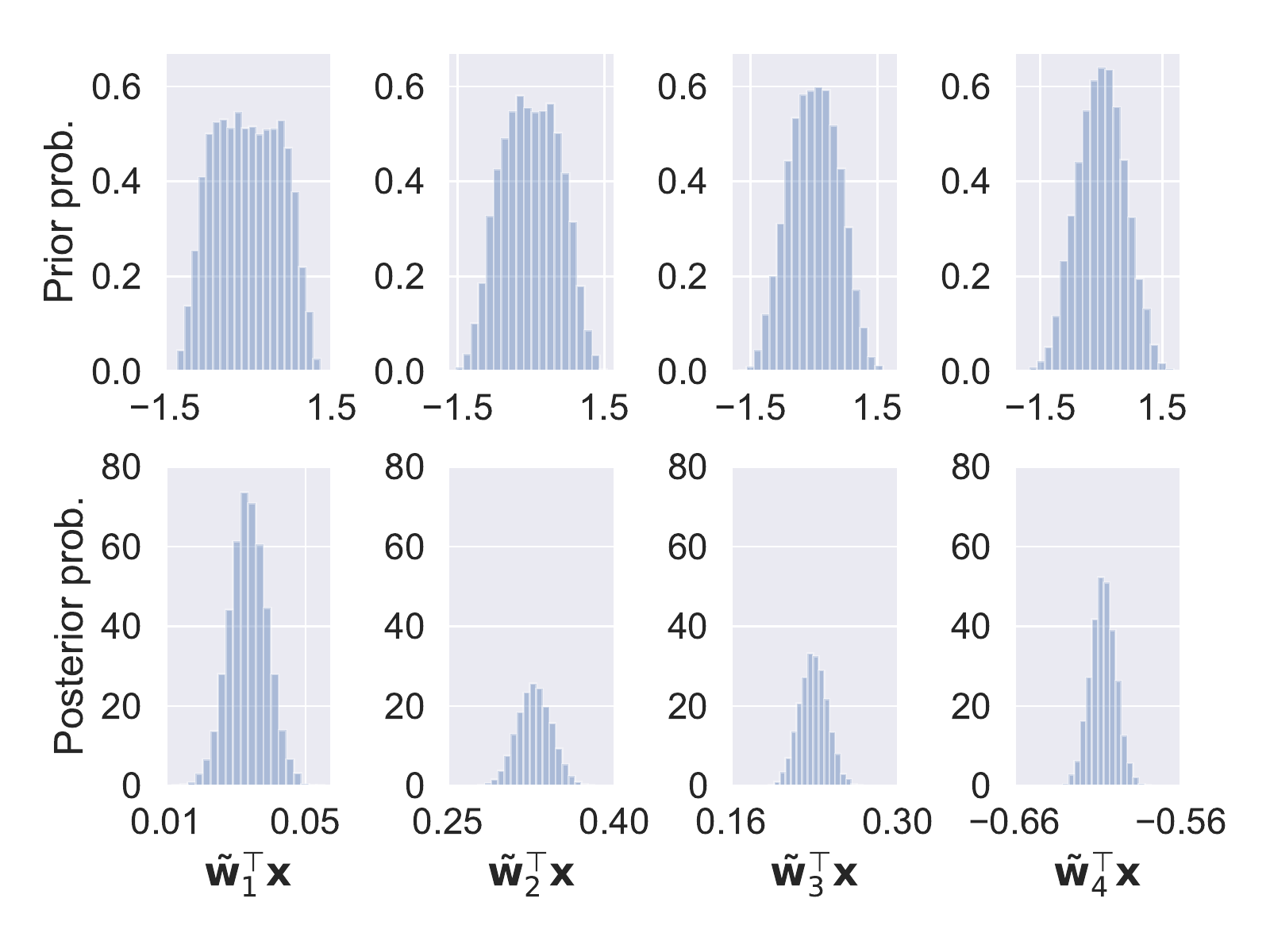}
	\caption{1D marginal prior and posterior distributions in the 4D active subspace.}
	\label{fig:marginals_actvar_4d_5}
\end{figure}

The Metropolis-Hastings algorithm described in Algorithm~\ref{alg:mcmc_as} is used to construct a Markov chain giving (correlated) posterior samples in the active subspace.
Its proposal variance was adjusted to $0.005$ in order to give an acceptance rate of $\approx 35\%$.
We compute 1,000,000 samples and regard the first 100,000 samples as part of the burn-in which are not considered as part of the final distribution.
The remaining samples give an effective sample size of about 12,000 which is enough to sufficiently represent a distribution in four dimensions.
The resulting distribution on the active variables, attained from the about 12,000 samples, is displayed in Fig.~\ref{fig:marginals_actvar_4d_5}.
Note that the $x$ scales of the upper and lower row of plots are quite different.
This due to the fact that if we did not change the lower $x$ scale, the lower histograms would basically become thin lines displaying no information about the variance of the distribution.
However, we see that the active variables are substantially informed which is exactly what we hoped to achieve.
Also note that the first plot in the upper row, displaying the 1D marginal prior distribution of the first active variable $\tilde{y}_1=\mathbf{\tilde{w}}_1^\top\mathbf{x}$, is almost a classical (rectangular) uniform distribution.
This is caused by the large contribution of only one parameter in $\mathbf{\tilde{w}}_1$ which is $\triangle k_{\text{is},(1,2)}$ in this case.
The more parameters contribute to an active variable, the more its marginal prior distribution differs from a rectangle, see $\tilde{y}_4 = \mathbf{\tilde{w}}_4^\top\mathbf{x}$ for example.

The samples in the inactive subspace $\mathbf{\tilde{z}}$ are computed as described in Subsection~\ref{ssec:mcmc_as} and composed with active posterior samples to give posterior samples in the original space.
Resulting 1D marginal statistics of the physical parameters are given in Table~\ref{tab:posterior} (left).
As expected, the physical parameters related to calibration parameters with significant components in the active subspace are highly informed.
The first calibration parameter ($\bar{k}^\text{log}_{\text{hyd},1}$) having a small but not negligible contribution according to the sensitivity values in Fig.~\ref{fig:eigvecs_4d_5_scores_5} e) is already only mildly informed.
The other parameters do not change or only very little because of the choice of the active subspace.

\begin{table}
	\centering
	\caption{Left: Posterior means and standard deviations of physical parameters.
		The informed parameters are highlighted in bold.
		Right: Highest 2D correlations for physical parameters.}
	\label{tab:posterior}
	\begin{tabular}{cccc}
		\hline\noalign{\smallskip}
		No. & Phys. par.          & Mean                           & Std. \\
		\hline\noalign{\smallskip}
		$1$ & $k_{\text{hyd},1}$  & $3.07 \times 10^2$             & $2.34 \times 10^2$ \\
		$2$ & $e_{\text{min},1}$  & $29.86$                        & $11.57$ \\
		$3$ & $e_{\text{max},1}$  & $44.49$                        & $12.90$ \\
		$4$ & $\alpha_1$          & $1.17$                         & $0.26$ \\
		$5$ & $k_{\text{is},1}$   & $\mathbf{5.18 \times 10^{-2}}$ & $\mathbf{3.98 \times 10^{-3}}$ \\
		$6$ & $k_{\text{sec},1}$  & $0.17$                         & $0.22$ \\
		$7$ & $e_{\text{sec},1}$  & $47.78$                        & $12.95$ \\
		\hline\noalign{\smallskip}
		$8$  & $k_{\text{hyd},2}$ & $70.62$                        & $55.81$ \\
		$9$  & $e_{\text{min},2}$ & $60.46$                        & $11.27$ \\
		$10$ & $e_{\text{max},2}$ & $1.20 \times 10^{2}$           & $16.14$ \\
		$11$ & $\alpha_2$         & $0.82$                         & $0.21$ \\
		$12$ & $k_{\text{is},2}$  & $\mathbf{4.52 \times 10^{-3}}$ & $\mathbf{1.61 \times 10^{-4}}$ \\
		$13$ & $k_{\text{sec},2}$ & $2.03 \times 10^{-2}$          & $3.23 \times 10^{-2}$ \\
		$14$ & $e_{\text{sec},2}$ & $1.76 \times 10^2$             & $25.99$ \\
		\hline\noalign{\smallskip}
		$15$ & $k_{\text{hyd},3}$ & $25.94$                        & $21.75$ \\
		$16$ & $e_{\text{min},3}$ & $95.71$                        & $14.18$ \\
		$17$ & $e_{\text{max},3}$ & $2.06 \times 10^2$             & $20.23$ \\
		$18$ & $\alpha_3$         & $0.43$                         & $0.14$ \\
		$19$ & $k_{\text{is},3}$  & $\mathbf{6.35 \times 10^{-4}}$ & $\mathbf{1.69 \times 10^{-5}}$ \\
		$20$ & $k_{\text{sec},3}$ & $6.21 \times 10^{-3}$          & $1.07 \times 10^{-2}$ \\
		$21$ & $e_{\text{sec},3}$ & $3.85 \times 10^2$             & $37.48$ \\
		\hline
	\end{tabular}
	\hfill
	\begin{tabular}{ccc}
		\hline\noalign{\smallskip}
		\multicolumn{2}{c}{Phys. par.} & Cor. coef. \\
		\hline\noalign{\smallskip}
		$e_{\text{min},1}$ & $e_{\text{max},1}$ & 0.89 \\
		$k_{\text{is},1}$  & $k_{\text{is},2}$  & 0.77 \\
		$e_{\text{min},3}$ & $e_{\text{max},3}$ & 0.70 \\
		$e_{\text{min},2}$ & $e_{\text{max},2}$ & 0.70 \\
		$k_{\text{hyd},2}$ & $k_{\text{sec},2}$ & 0.66 \\
		$k_{\text{is},1}$  & $k_{\text{hyd},2}$ & 0.64 \\
		$k_{\text{is},1}$  & $k_{\text{sec},2}$ & 0.63 \\
		$k_{\text{hyd},2}$ & $k_{\text{is},2}$  & 0.59 \\
		$k_{\text{sec},1}$ & $k_{\text{is},3}$  & 0.57 \\
		$k_{\text{sec},2}$ & $k_{\text{sec},3}$ & 0.56 \\
		$k_{\text{is},2}$  & $k_{\text{sec},2}$ & 0.52 \\
		\hline
	\end{tabular}
\end{table}

Additionally, Table~\ref{tab:posterior} (right) displays the highest resulting two-dimensional posterior correlation coefficients of the physical parameters in LuKARS.
They are consistent with the components of corresponding calibration parameters that show up in the 4D active subspace.
The largest correlation occurs between $e_{\text{min},1}$ and $e_{\text{max},1}$.
A large correlation coefficient of $0.7$ is also found between the respective storage values of Hyd~2 and Hyd~3, $e_{\text{min},2}$ and $e_{\text{max},2}$, as well as between $e_{\text{min},3}$ and $e_{\text{max},3}$.

\begin{figure}
	\centering
	\includegraphics[width=\textwidth]{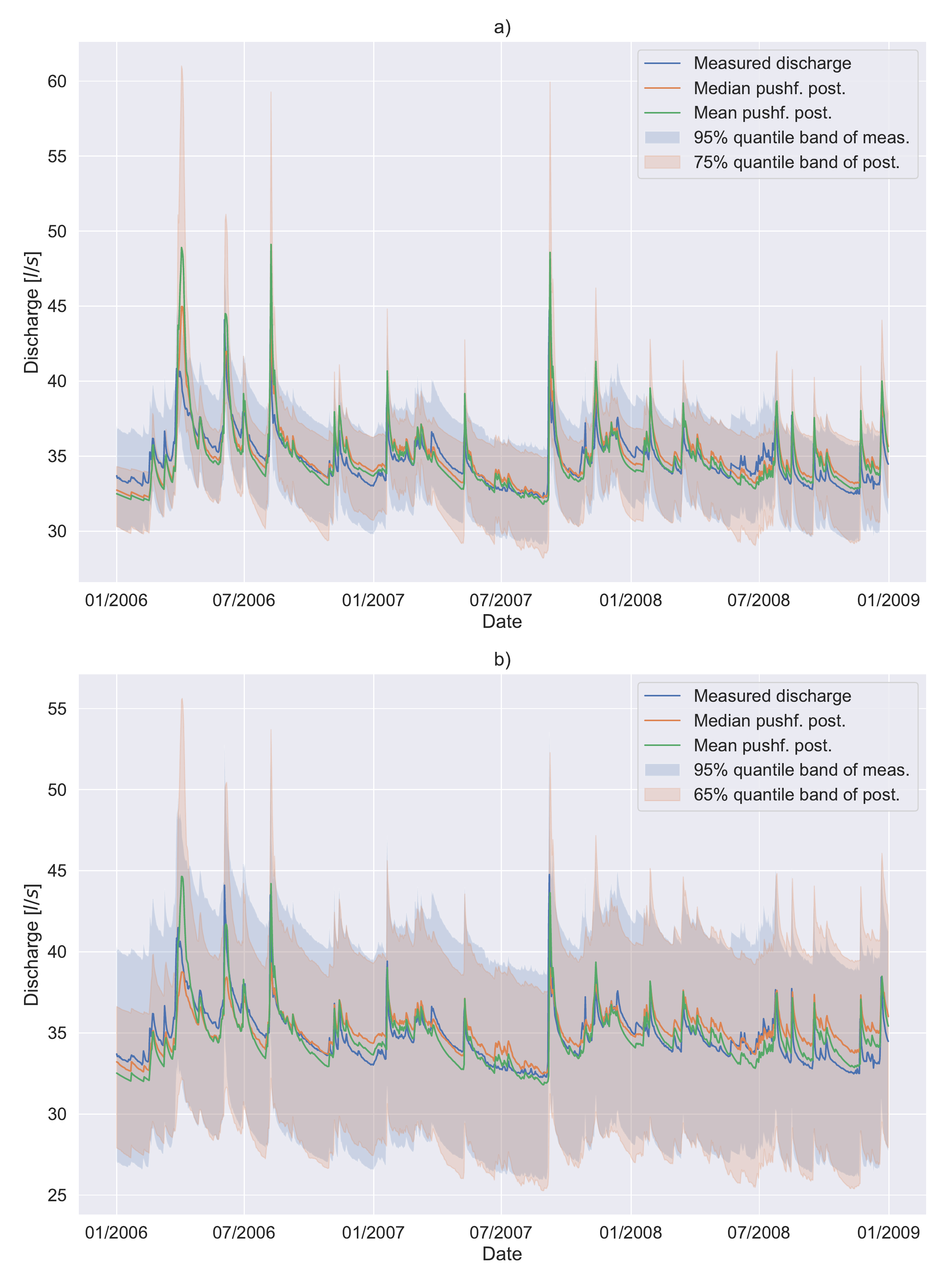}
	\caption{Push-forward distributions of the posteriors gained with a 4D (a) and 1D (b) subspace, assuming a 5\% and 10\% noise level, respectively, along with their mean and median.}
	\label{fig:quantiles}
\end{figure}

As a verification for our choice of a 4D subspace, we show that the uncertainty in the approximated posterior distribution is dominated by the uncertainty in the data, and not by approximation errors caused by dimension reduction.
We do this by approximating the posterior's \textit{push-forward distribution}, i.\,e., the distribution gained by propagating the approximated posterior through the parameter-to-observation operator $\mathcal{G}$, which models the discharge values for a given input parameter.
Hence, we computed 1,000 samples of the distribution $\mu_{\text{post}}(\{\mathbf{x} \,|\, \mathcal{G}(\mathbf{x})\in\cdot\})$, where $\mu_\text{post}$ denotes the posterior distribution with density $\rho_{\text{post}}$.
Fig.~\ref{fig:quantiles}~a) shows the 95\% quantile band around the data with 5\% additive Gaussian noise assumed together with a 75\% quantile band around the median of the posterior's push-forward distribution.
More loosely speaking, the plot shows that around 75\% of discharges simulated with random parameters drawn from the posterior will lie within the inherent uncertainty of the observed discharge.
The uncertainty in the dynamics around the measured discharges in the Kerschbaum spring is matched well by the uncertainty of the push-forward posterior distribution which confirms the choice of a 4D subspace.
Since we started with an uninformed prior (a uniform distribution), we can not expect to end up with a push-forward posterior much more certain than the uncertainty in the experiments.
At this point, we would like to emphasize that it is possible to get a reasonably good approximation of the posterior by considering only 4 directions in the space of 21 parameters.
In this manner, particularly regarding the low flow conditions and the recession limbs of the peak discharges, we can observe a variation of only up to $\pm$5l/s which is roughly the variation due to experimental noise.
Also, the mean and the median of the push-forward distribution give results agreeing with the data which, in addition, supports the decision for a 4D subspace.

Additionally, this type of plot shows that different decisions for the dimension of the active subspace lead to different posterior approximations.
Fig.~\ref{fig:quantiles}~b) shows the push-forward distribution of a posterior gained with a 1D subspace.
However, interestingly, assuming a noise level of 10\% and taking only a 1D subspace also leads to usable results in the sense that the corresponding approximation to the posterior's push-forward distribution matches the inherent uncertainty in the data well.
However, note that the $r^2$ score of $\tilde{G}_\mathbf{d}$ was only about $0.23$ in this case.
Although we see that the assumptions are rather unrealistic and the approximation quality of $\tilde{G}_\mathbf{d}$ is too bad, it is worth noting that already the first eigenvector contains some information about the posterior meaning that the mean and median of the corresponding push-forward posterior give reasonable discharge values in comparison with the data as shown in Fig.~\ref{fig:quantiles}~b).

%% file: discussion.tex
\section{Discussion}
\label{sec:discussion}
Besides the introduction of the active subspace method as a technique for dimension reduction in Bayesian inverse problems to the karst hydrology community, a major aim of our work, as mentioned in Section~\ref{sec:introduction}, was to perform a parameter inference in the Bayesian sense providing information about the behavior of our model and its uncertainties.
For the LuKARS model of the Kerschbaum spring in Waidhofen a.d. Ybbs, we found a 4-dimensional subspace of the original 21-dimensional parameter space.
This does, however, not mean that only 4 individual physical model parameters are informed by the discharge data since an active subspace represents a linear combination of sensitive parameters, represented by the parameters in the eigenvectors corresponding to dominant eigenvalues.
In this regard, the relation between sensitive parameters in each dimension of the active subspace provides deeper insights into the model behavior than just the sensitivities of individual parameters.
The results, from a broader perspective, show that 7 physical/model parameters are most sensitive.
These parameters consist of coefficients for the baseflow storages ($k_{\text{is},i}$, $i=1,2,3$), for the quickflow storages ($k_{\text{hyd},i}$, $i=1,2$), and for the secondary spring discharges ($k_{\text{sec},i}$, $i=1,2$).

The remaining paragraphs in this section are devoted to give a detailed hydrological interpretation of the results showed in the previous section.
These interpretations are based on the following information.
The observed spring discharge is modeled as the sum of the relative contribution of each hydrotope.
Moreover, the LuKARS model of the Kerschbaum spring has fast responding hydrotopes (i.e., hydrotopes that quickly deliver water to the karst spring after precipitation events, e.g., Hyd~1) and slow responding hydrotopes (i.e., hydrotopes, which slowly deliver water to the spring after precipitation events, e.g., Hyd~3).

Parameters 5, 12, and 19 show the largest contributions in the first 4 eigenvectors in Fig.~\ref{fig:eigvecs_4d_5_scores_5}~a)-d).
These parameters correspond to the $k_{\text{is}}$ physical parameters, which delimit the flow contributions from the hydrotopes to the linear baseflow storage.
As derived in \citeA{bittner2018modeling}, the baseflow storage exhibits a relatively constant discharge behavior with a small temporal variability and its discharge coefficient $k_{\text{b}}$ was not changed within the presented research study.
Since the outflow from the baseflow storage is controlled by its variable storage ($e_{\text{b}}$) and its constant discharge coefficient ($k_{\text{b}}$), the hydrotope discharge coefficients for the groundwater recharge ($k_{\text{is}}$) also affect the baseflow discharge and its temporal dynamics since they control $e_{\text{b}}$.
Given that $k_{\text{b}}$ was not included as a calibration parameter, the $k_{\text{is}}$ parameters are responsible to maintain the baseflow contribution as derived by \citeA{bittner2018modeling} and are most informed in the first eigenvector when applying the active subspace method.

Although parameters 5, 12, and 19 have the same physical interpretation, we can observe that they display different sensitivities for the different hydrotopes.
This is due to the fact that different hydrotopes cover areas which are different in extension (Hyd~1 - 13\%, Hyd~2 - 56\% and Hyd~3 - 27\%).
Therefore, the interpretation of the most important parameters occurring in an eigenvector, should both consider the physical meaning of the parameter and the relative contribution of each single hydrotope to the total spring discharge, which is highly affected by the relative area covered by the hydrotope.
In this specific case, parameter 12, associated with Hyd~2, displays the largest value since it covers the largest area in the Kerschbaum spring catchment and thus has a significant contribution to the total spring discharge.
Parameter 5 has the second largest value although Hyd~1 ranks as third in terms of coverage area.
This is explained by the fact that Hyd~1 provides the most dynamic and variable discharge behavior of all hydrotopes.
Hence, the discharge contribution from Hyd~1 is essential to reproduce the discharge dynamics observed in the Kerschbaum spring.
Hyd~3 has the smallest contribution in eigenvector 1, which can be explained by its more constant and less variable discharge behavior as compared to Hyd~1 and its smaller spatial share as compared to Hyd~2.
Hence, although Hyd~3 has a larger area covered than Hyd~1, parameter 19 is less dominant than parameter 5.

Although parameters~1 and 8 (involving $k_{\text{hyd}}$ values of Hyd~1 and Hyd~2) do not show up in eigenvector~1, their contribution to eigenvectors~2 to 4 is worth discussing.
These parameters follow a different ranking as compared to the $k_{\text{is}}$ parameters, suggesting a larger sensitivity of parameter~1 from Hyd~1 as compared to parameter~8 from Hyd~2.
Since the $k_{\text{hyd}}$ parameters constrain the quickflow dynamics originating from each hydrotope, we argue that this ranking is the result of the different hydrological behaviors each hydrotope is supposed to simulate.
Considering that Hyd~1, which shows the most dynamic behavior in response to precipitation or melt events, has a large contribution to the temporal variability of the discharge in the Kerschbaum spring, the importance of adequately representing the quickflow dynamics from Hyd~1 can be regarded as more important than the relative space covered by each hydrotope.

It is interesting to observe that the posterior means of the informed physical model parameters ($k_{\text{is},i}$, $i=1,2,3$, see Table~\ref{tab:posterior}) are close to corresponding calibrated parameters found by \citeA{bittner2018modeling}.
Moreover, the standard deviations of the $k_{\text{is}}$ parameters in the posterior distribution are smaller as compared to the standard deviations found for the posterior distributions of all other physical parameters.
This provides evidence that we can use the components of the first eigenvectors derived from the active subspace method to show which parameters get individually updated from the prior to the posterior distribution.
Moreover, in comparison to the parameters obtained by manual calibration, we additionally obtain an uncertainty specification related to any model parameter.

It is striking that the correlations between $e_{\text{min},i}$ and $e_{\text{max},i}$ (for each $i=1,2,3$) are high although they are not very dominant according to the first four eigenvectors.
The reason for this is that this correlation is already present in the prior distribution on the physical parameters (see Eq.~\eqref{eq:e_minmax}).
However, we argue that the higher positive correlation coefficient between $e_{\text{min,1}}$ and $e_{\text{max,1}}$ results from the dependence of the overall model output on the quickflow dynamics of Hyd~1 \cite{bittner2018modeling}.
The dynamics of the quickflow depend strongly on the difference between $e_{\text{min},i}$ and $e_{\text{max},i}$ (see Eq.~\eqref{eq_hyd_discharge}) and thus results in a high correlation coefficient between both storage thresholds.
The positive correlations between the discharge coefficients of Hyd~2, in particular between $k_{\text{hyd},2}$ and $k_{\text{sec},2}$, $k_{\text{hyd},2}$ and $k_{\text{is},2}$ as well as between $k_{\text{is},2}$ and $k_{\text{sec},2}$, highlight the strong interdependence of all discharge components that originate from Hyd~2.
The strongest correlation (0.66) is between the discharge coefficient of the quickflow ($k_{\text{hyd},2}$) and the discharge coefficient of the secondary spring discharge ($k_{\text{sec},2}$).
This means that, if we increase $k_{\text{hyd},2}$ and not $k_{\text{sec},2}$, the quickflow contribution increases disproportionately and the total simulated spring discharge would overestimate the observed peak discharges.
The same relationship holds for the strong correlation (0.59) between the discharge coefficient of the quickflow ($k_{\text{hyd},2}$) and the discharge parameter of the groundwater recharge ($k_{\text{is},2}$) as well as between the $k_{\text{is},2}$ and $k_{\text{sec},2}$ (0.52).
These correlations confirm the fact that if we increase the discharge coefficient of one discharge component in a certain hydrotope, we need to simultaneously increase all other discharge coefficients in the same hydrotope to get a similar model output.
If the other coefficients were not changed accordingly, we would disproportionately increase one discharge component (e.\,g., quickflow from Hyd~2) relative to others (e.\,g., $k_{\text{is},2}$ or $k_{\text{sec},2}$); so, the hydrotope would show a different hydrological behavior.
This highlights that the parameter dependencies within each hydrotope individually help to maintain the hydrological behavior that is typical for each hydrotope.
The reason why only the discharge parameters of Hyd~2 show high correlation coefficients is that Hyd~2 covers more than 50\% of the Kerschbaum spring recharge area and, thus, has the highest contributions to the total spring discharge.
The correlations between various discharge coefficients of different hydrotopes, e.\,g., $k_{\text{is,1}}$ and $k_{\text{is,2}}$, are interpreted as a consequence of the parameter constraints introduced in Eq.~\eqref{eq:constraints}, similar to the dependence between $e_{\text{min},i}$ and $e_{\text{max},i}$ values.

We conclude by commenting on the limits and transferability of the active subspace method.
One of the major disadvantages of the method is the need for gradients in the identification of the active subspace.
Computing gradient information can be computationally expensive, especially if there are no alternatives to using a finite difference approach as, e.\,g., adjoint formulations \cite{plessix2006review}.
In the given case, the computational costs for using central finite differences are reasonable since the model is sufficiently cheap.
However, for more expensive models such an approach can be intractable.
There are recent advances for computing active subspaces in a derivative-free way \cite{tripathy2016gaussian}, however, the computation of derivatives is replaced by a non-trivial non-convex optimization problem.
Since the framework of active subspaces is quite general and formulated without too many restricting assumptions, we consider its application as highly transferable.
This claim is supported by several applications of the method for complex physical models, e.\,g., \citeA{erdal2019global,constantine2015exploiting,constantine2017time,diaz2018modified,jefferson2015active}.
The proposed use of the active subspace method extends the available tools for parameter and uncertainty estimation in hydrology.
A comparison among all available methods is out of the scope of this work and should be pursued in the future through a collaborative work in the community.
We hence provide only a brief comparison with DREAM, described in \citeA{vrugt2016markov} and used to accelerate MCMC in \citeA{vrugt2009accelerating}, as it has found wide recognition in Bayesian analysis of hydrological models.
The main differences of DREAM compared to the presented approach are two-fold.
First, basic DREAM runs multiple Markov chains in parallel, exchanging chain elements in a way retaining favorable properties like ergodicity and the Markov property (memorylessness).
Secondly, a randomized subspace sampling strategy is then used to avoid inefficient mixing of Markov chains in higher dimensions.
Our proposed approach, however, does not need an additional sampling strategy, since the Markov chains are moving in the dominant low-dimensional subspace, whereas for DREAM the chain is still evolving in the full-dimensional space.

%% file: summary.tex
\section{Summary}
\label{sec:summary}
This manuscript shows results from a parameter study of a karst aquifer model for the Kerschbaum spring recharge area.
The model uses 21 parameters to simulate the discharge behavior of the Kerschbaum karst spring in Waidhofen a.d. Ybbs.
The study consists of a parameter inference in the Bayesian sense and a (global) sensitivity analysis.
Since these problems have a non-trivial dimension, we first check for low-dimensional structure, if present, hidden in the inference process and exploit the so-called \textit{active subspace method} for this.
Additionally, without further expensive computations, we are then able to derive global sensitivity metrics.

It seems that the inference process is indeed intrinsically low-dimensional.
Although the LuKARS model for the Kerschbaum spring has 21~calibration parameters, given the parameter constraints in Eq.~\eqref{eq:constraints}, we find its dominant parameters and obtain well-constrained values for them by means of Bayesian inversion in the identified active subspace.
In particular, we decide to reduce the Bayesian inverse problem from a 21D to a 4D problem which is verified by showing that the push-forward distribution of the approximated posterior has a promising similarity with the uncertainty in the data.
The 1D and 2D posterior statistics, which differ a lot from corresponding prior statistics for dominant parameters, are computed to quantify uncertainty in the inference caused by measurement errors in the data.

Eventually, the active subspace method shows again to be valuable for Bayesian inference and sensitivity analysis in complex high-dimensional problems.
The results are, however, rather useful from a computational perspective.
The in-depth validation of the model with further sensitivity analyses, more interesting from a hydrological perspective, and a discussion of the consequences for the community are out of scope and, hence, not part of this study, but will follow in future research.
In particular, we want to investigate the hydrological features that lead to the present dimensional reduction.

%% file: appendix.tex
\section{Model equations}
\label{app:model}
In LuKARS, the following balance equation is solved for each individual hydrotope:

\begin{equation}
\label{eq_hyd_balance}
\frac{de_i}{dt}=\begin{cases}
S_i-\frac{Q_{\text{sec},i}+Q_{\text{is},i}+Q_{\text{hyd},i}}{a_i} 
& \text{if $e_i>0$}\\
0 & \text{if $e_i=0$}
\end{cases}
\end{equation}
\(e_i\) is the water level [L] in hydrotope \(i\), \(t\) [T] indicates the time and \(S_i\) is a hydrotope-specific sink and source term in form of a mass balance of precipitation, snow melt, evapotranspiration and interception.
We used the temperature index approach from \citeA{martinec1960degree} to calculate snow melt.
Interception was estimated based on indications for beech forests in \citeA{dvwk1996}.
Then, evapotranspiration was calculated based on the method of \citeA{thornthwaite1948approach}.
\(Q_{\text{sec}}\) [L$^3$T$^{-1}$] summarizes all flow terms that do not contribute to the discharge at an investigated karst spring, i.\,e., secondary spring discharge and overland flow.
\(Q_{\text{is},i}\) [L$^3$T$^{-1}$] represents the discharge from hydrotope \(i\) to a linear baseflow storage, considered as groundwater recharge. \(Q_{\text{hyd},i}\) [L$^3$T$^{-1}$] is a hydrotope-specific quickflow component through preferential flow paths (e.\,g., subsurface conduits) with a direct connection to the spring outlet. \(a_i\) [L$^2$] is the space covered by a respective hydrotope.

The following balance equation is solved for the baseflow storage:

\begin{equation}
\label{eq_baseflow_balance}
\frac{de_\text{b}}{dt}=\begin{cases}
\frac{\Sigma(Q_{\text{is},i})-Q_\text{b}}{A}
& \text{if ${e_\text{b}}>0$}\\
0 & \text{if ${e_\text{b}}=0$,}
\end{cases}
\end{equation}
\(e_\text{b}\) is the water level [L] in the baseflow storage and \(\Sigma(Q_{\text{is},i})\) [L$^3$T$^{-1}$] integrates the flows from all hydrotopes to the baseflow storage.
\(Q_\text{b}\) [L$^3$T$^{-1}$] indicates water flow from the storage B to the spring and simulates the matrix contribution from the saturated zone to the spring discharge.
The variable \(A\) [L$^2$] is the space of the entire recharge area. The discretized forms of \ref{eq_hyd_balance} and \ref{eq_baseflow_balance}, as shown in \ref{eq_hyd_discr} and \ref{eq_baseflow_discr}, are solved for each time step \(n\):

\begin{equation}
\label{eq_hyd_discr}
e_{i,n+1}=\max\left[0,e_{i,n}+\left(S_{i,n}-\frac{Q_{\text{sec},i,n}+Q_{\text{is},i,n}+Q_{\text{hyd},i,n}}{a_i}\right)*\Delta t\right]
\end{equation}

\begin{equation}
\label{eq_baseflow_discr}
e_{\text{b},n+1}=\max\left[0,e_{\text{b},n}+\left(\frac{\Sigma(Q_{\text{is},i,n})-Q_{\text{b},n}}{A}\right)*\Delta t\right] 
\end{equation}

The discharge terms are computed as:

\begin{equation}
\label{eq_hyd_discharge}
Q_{\text{hyd},i,n}=\varepsilon \left[\frac{\max(0,e_{i,n}-e_{\text{min},i})}{e_{\text{max},i}-e_{\text{min},i}}\right]^{\alpha_i} *\frac{k_{\text{hyd},i}}{l_{\text{hyd},i}}*a_i
\end{equation}

\begin{equation}
\label{eq_is_discharge}
Q_{\text{is},i,n}=k_{\text{is},i}*e_{i,n}*a_i
\end{equation}

\begin{equation}
\label{eq_sec_discharge}
Q_{\text{sec},i,n}=k_{\text{sec},i}*\max(0,e_{i,n}-e_{\text{sec},i})*a_i
\end{equation}

\begin{equation}
\label{eq_baseflow_discharge}
Q_{\text{b},n}=k_\text{b}*e_{\text{b},n}*A
\end{equation}

\(e_{\text{max},i}\) [L] and \(e_{\text{min},i}\) [L] are the upper and lower storage thresholds of hydrotope~\(i\).
The exponent \(\alpha_i\) controls the magnitude of the quickflow component from each hydrotope.
\(e_{\text{sec},i}\) [L] represents a hydrotope-specific activation level for \(Q_{\text{sec}}\).
\(k_{\text{is},i}\) [LT$^{-1}$] and \(k_{\text{sec},i}\) [LT$^{-1}$] are the specific discharge parameters for \(Q_{\text{is},i}\) [L$^3$T$^{-1}$] and \(Q_{\text{sec},i}\) [L$^3$T$^{-1}$].
\(k_{\text{hyd},i}\) [L$^2$T$^{-1}$] indicates the specific discharge parameter for the quickflow and \(l_{\text{hyd},i}\) [L] is the mean distance of hydrotope \(i\) to the adjacent spring, allowing to account for the relative location and distribution of hydrotope \(i\) in a specific recharge area.
The ratio between \(k_{\text{hyd},i}\) and \(l_{\text{hyd},i}\) is the hydrotope discharge coefficient.
Then, the dimensionless connectivity/activation indicator $\varepsilon$ is defined as:

\begin{equation}
\label{eps_1}
\varepsilon_{n+1}=0 \quad \text{if} \; \begin{cases}
\text{$\varepsilon_n=0$ \& $e_{i,n+1}<e_{\text{max},i}$ or}\\
\text{$\varepsilon_n=1$ \& $e_{i,n+1} \leq e_{\text{min},i}$}
\end{cases}
\end{equation}

\begin{equation}
\label{eps_2}
\varepsilon_{n+1}=1 \quad \text{if} \; \begin{cases}
\text{$\varepsilon_n=0$ \& $e_{i,n+1} \geq e_{\text{max},i}$ or}\\
\text{$\varepsilon_n=1$ \& $e_{i,n+1}>e_{\text{min},i}$}.
\end{cases}
\end{equation}

%% file: acknowledgments.tex
This collaborative research is a result of the \textit{UNMIX} project (\textit{UNcertainties due to boundary conditions in predicting MIXing in groundwater}), which is supported by \textit{Deutsche Forschungsgemeinschaft} (DFG) through \textit{TUM International Graduate School for Science and Engineering} (IGSSE), GSC 81.
Additional financial support for BW, SM, and MTP was provided by the German Research Foundation (DFG, Project WO 671/11-1).
GC also acknowledges the support of the Stiftungsfonds f\"ur Umwelt\"okonomie und Nachhaltigkeit GmbH (SUN).
The authors further thank the water works Waidhofen a.d. Ybbs for providing the relevant spatial and time series data.
For reproducibility, code and data are available at \path{https://bitbucket.org/m-parente/2019-karst-unmix-data/}.